\documentclass{jfm}
\usepackage{amssymb, bm,amsbsy,times,amsmath}
\usepackage{graphicx}
\usepackage{epstopdf, epsfig}
\usepackage{appendix}
\pdfoutput=1

\usepackage{xcolor}

\shorttitle{Fully developed anelastic convection with no-slip boundaries}
\shortauthor{C. A. Jones, K. A. Mizerski and M. Kessar }

\title{Fully developed anelastic convection with no-slip boundaries}

\author{Chris A. Jones\aff{1}
  \corresp{\email{cajones@maths.leeds.ac.uk}},
  Krzysztof A. Mizerski\aff{2}
 \and Mouloud Kessar\aff{3}}

\affiliation{\aff{1}Department of Applied Mathematics, University of Leeds,
Leeds, LS2 9JT, UK
\aff{2} Department of Magnetism, Institute of Geophysics, Polish Academy of Sciences, ul. Ksiecia Janusza 64, 01-452 Warsaw, Poland
\aff{3}Universit\'e de Paris, Institut de physique du globe de Paris, CNRS, IGN, F-75005 Paris, France}

\begin{document}

\maketitle

{\center{Manuscript at \today}} 
\medskip

\begin{abstract}
Anelastic convection at high Rayleigh number in a plane parallel layer with no slip boundaries is considered.
Energy and entropy balance equations are derived, and they are used to develop scaling laws for the heat transport and the Reynolds number. 
The appearance of an entropy structure consisting of a well-mixed uniform interior, bounded by thin
layers with entropy jumps across them, makes it possible to derive explicit forms for these scaling laws.  These are given in terms of
the Rayleigh number, the  Prandtl number, and the bottom to top  temperature ratio, which measures how stratified the
layer is. The top  and bottom boundary layers are examined and they are found to be very different, unlike in the Boussinesq case. 
Elucidating the structure of these boundary layers plays a crucial part in determining the scaling laws.
Physical arguments governing  these boundary layers are presented, concentrating on the case
in which the boundary layers are thin even when the stratification is large, the incompressible boundary layer case. Different
scaling laws are found, depending on whether  the viscous dissipation is primarily in the boundary layers or in the bulk.
The cases of both  high and low Prandtl number are considered. Numerical simulations of no-slip anelastic convection up to a Rayleigh number of $10^7$ have been performed and our theoretical predictions are compared with the numerical results.  
 
\end{abstract}

%\begin{keywords}
%Authors should not enter keywords on the manuscript, as these must be chosen by the author during the online submission process and will then be added during the typesetting process (see http://journals.cambridge.org/data/\linebreak[3]relatedlink/jfm-\linebreak[3]keywords.pdf for the full list)
%\end{keywords}

\section{Introduction} %%%%

The problem of the influence of density stratification on developed convection is of great
importance from the astrophysical point of view. Giant planets and stars are typically strongly
stratified and the anelastic approximation, see e.g. Ogura \& Phillips (1962), Gough (1969) and Lantz \&
Fan (1999), is commonly used to describe convection in their interiors,  e.g. Toomre \textit{et al.} (1976), Glatzmaier \& Roberts (1995), Brun \& Toomre (2002), Browning \textit{et al.} (2006), Miesch \textit{et al.} (2008), 
Jones \& Kuzanyan (2009), Jones \textit{et al.} (2011),  Verhoeven \textit{et al.} (2015), Kessar \textit{et al.} (2019) and many others. 
The anelastic approximation is based on the convective system being a small
departure from the adiabatic state, which is appropriate for large scale systems with developed,
turbulent and thus strongly-mixing convective regions. The small departure from adiabaticity
induces convective velocities much smaller than the speed of sound, so sound waves are neglected
in the dynamics. We consider a plane layer of fluid between two parallel plates distance $d$ apart,
and we assume the turbulent flow is spatially homogeneous in the horizontal direction. We also
assume that the convection is in a statistically steady state, so that the time-averages of time-derivative
terms can be neglected. Most astrophysical applications are in spherical geometry, but
the simpler plane layer problem is a natural place to start our investigation of high Rayleigh
number anelastic convection. 

In convection theory, we try to determine the dependencies of
the superadiabatic heat flux and the convective velocities (measured by the Nusselt, $Nu$, and
Reynolds $Re$ numbers) on the driving force measured by the imposed temperature difference between the top and bottom plates, 
i.e. on the Rayleigh number, $Ra$, and on the Prandtl number,
$Pr$ (the ratio of the fluid kinematic viscosity to its thermal diffusivity). Here we aim to discover how these
dependencies are affected by the stratification. We rely strongly on the theory of Grossmann \&
Lohse (2000) (further developed later and updated in Stevens et al. 2013) developed for Boussinesq,
i.e. non-stratified, convection. However, compressible convection differs strongly from the
Boussinesq case, with the latter mostly corresponding to experimental situations. There are two
crucial differences, which have very important consequences for the dynamics of convection.
Firstly, in the compressible case the viscous heating and the work of the buoyancy force are no
longer negligible compared to the heat transport. Secondly, in stratified convection the boundary
layers and the flow velocities are different at the top of the layer and the bottom of the layer,
unlike the Boussinesq case where the top and bottom boundary layers have the same structure
and the temperature of the well-mixed interior is exactly half way between the temperature of the
top and bottom plates. So although our approach is based on that of Grossmann \& Lohse (2000),
there are additional novel features required in the compressible convection case. We develop the
theory of fully developed convection in stratified systems and study the dependence of the total
superadiabatic heat flux and the amplitude of convective flow on the number of density scale
heights in the layer. The scaling laws, i.e. the dependencies of the Nusselt and Reynolds numbers
on the Rayleigh and Prandtl numbers are the same as in the Boussinesq convection.

In this paper, we concentrate on the convective regimes which seem to be
most relevant to current numerical capabilities, i.e. regimes most easily achieved by numerical
experiments. These are the regimes where the thermal dissipation is dominated by the thermal
boundary layer contribution. It is those regimes, denoted by $I_u$, $I_l$, $II_u$ and $II_l$ on the phase
diagram, figure 2 of Grossmann \& Lohse (2000), that in the Boussinesq case correspond to
Rayleigh numbers less than about 1012. It must be noted, however, that contrary to the Boussinesq
case, which is well established by numerous experimental and numerical investigations, there
are to date no experiments on fully turbulent stratified convection, due to the difficulties of achieving significant
density stratification in laboratory settings. Some experiments are being developed using the
centrifugal acceleration in rapid rotating systems to enhance the effective gravity (Menaut {\it{et al.}} 2019). 
There have also been some numerical investigations of anelastic convection in a plane layer, mostly focussed on
elucidating how well the anelastic approximation performs compared to fully compressible convection, Verhoeven {\it et al.} (2015)
and  Curbelo {\it et al.} (2019). This latter paper notes that the top and bottom boundary layer structures that occur in the case of high Prandtl number
are different.

In addition
to the dependence on Rayleigh and Prandtl number, our problem depends on how stratified the
layer is, which can be estimated by the ratio $\Gamma$ of the temperature at the bottom of the the layer
$T_B$ to the temperature at the top $T_T$. When $\Gamma$ is close to unity the layer is nearly Boussinesq, but
when $\Gamma$ is large there are many temperature and density scale heights within the layer. We aim to
derive the functional form of $Nu(\Gamma,\,Ra,\,Pr)$ and $Re(\Gamma,\,Ra,\,Pr)$, but we cannot derive reliable
numerical values for the prefactors in anelastic convection. Since experiments are not available,
this will require high resolution high Ra simulations, which are being developed, but are not yet
in a state to determine the prefactors accurately.

In \S\,2 we derive the anelastic equations and the reference states we use, and outline the structure of high Rayleigh number 
anelastic convection. Further details of the form of the anrelastic temperature perturbation are given in appendix A. 
In \S\,3 we derive the energy and entropy production integrals, which are the fundamental building blocks
for developing convective scaling laws. In sections \S\,4 and \S\,5 we derive the physical arguments used to get the key properties
of the top and bottom boundary layers. In \S\,6 we derive the scaling laws for the case where the viscous dissipation is primarily in the
boundary layers. The case where the dissipation is mainly in the bulk is dealt with in appendix B. \S\,7 gives the results of our numerical simulations,
comparing them with our theoretical results. Our conclusions are in \S\,8.

\section{Fully developed compressible convection under the anelastic approximation}
Consider a layer of compressible perfect gas between two parallel plates,
of thickness $d$, periodic in horizontal directions, the evolution
of which is described by the set of the Navier-Stokes, mass conservation
and energy equations under the anelastic approximation,
\begin{equation}
\frac{\partial\mathbf{u}}{\partial t}+\left(\mathbf{u}\cdot\nabla\right)\mathbf{u}=-\nabla\left(\frac{p}{\bar{\varrho}}\right)+\frac{g}{c_{p}}s\hat{\mathbf{e}}_{z}+\frac{\mu}{\bar{\rho}}\left[\nabla^{2}\mathbf{u}+\frac{1}{3}\nabla\left(\nabla\cdot\mathbf{u}\right)\right],\label{eq:2_1}
\end{equation}
\begin{equation}
\nabla\cdot\left(\bar{\varrho}\mathbf{u}\right)=0,\label{eq:2_2}
\end{equation}
\begin{equation}
\bar{\rho}\bar{T}\left[\frac{\partial s}{\partial t}+\mathbf{u}\cdot\nabla s \right]=k\nabla^{2}T+\mu\left[q+\partial_{j}u_{i}\partial_{i}u_{j}-\left(\nabla\cdot\mathbf{u}\right)^{2}\right],\label{eq:2_3}
\end{equation}
\begin{equation*}
\refstepcounter{equation}
\frac{p}{\bar{p}}=\frac{\rho}{\bar{\rho}}+\frac{T}{\bar{T}},\qquad s=c_{v}\frac{p}{\bar{p}}-c_{p}\frac{\rho}{\bar{\rho}},
\quad \gamma= \frac{c_p}{c_v}, \quad c_p-c_v=R,
\eqno{(\theequation \textit{a,b,c,d})} \label{eq:2_4}
\end{equation*}
where
\begin{equation}
q=(\partial_{j}u_{i})(\partial_{j}u_{i})+\frac{1}{3}\left(\nabla\cdot\mathbf{u}\right)^{2},\label{eq:2_5}
\end{equation}
$\mathbf{u}$ being the fluid velocity, $p$ the pressure, $\rho$ the density, $T$ the temperature and $s$ the entropy.
Barred variables are adiabatic reference state variables, unbarred variables denote the perturbation from the reference state due
to convection.  The dynamic viscosity $\mu=\bar{\rho}\nu$, the thermal conductivity
$k$, gravity $g$ and the specific heats at constant pressure, $c_p$, and constant volume, $c_v$, are all assumed constant. The bounding plates
are no-slip and impenetrable, so $\bf u = 0$ there.
We consider the constant entropy boundary conditions 
\begin{equation}
s= \Delta S \quad \textrm{at} \quad  z=0, \qquad s = 0   \quad \textrm{at} \quad z=d.
\label{eq:2_6}
\end{equation}
 Note that we do not replace the thermal 
diffusion term in (\ref{eq:2_3}) by
an entropy diffusion term, as is often done in anelastic approaches. 
With our no-slip boundaries, there will be boundary layers which may be laminar even at very high
Rayleigh number, and entropy diffusion is not appropriate if laminar boundary layers are present.
We discuss the additional
issues raised by adopting constant temperature boundary conditions rather than constant entropy conditions in Appendix A.

In the anelastic approximation,  the full variables, $\hat p$, $\hat \rho$ and $\hat T$, 
are expanded in terms of the small parameter $\epsilon$, which is defined precisely in equation (\ref{eq:2_10}) below,  so 
\begin{equation} 
\hat p = \bar p + \epsilon  p, \quad \hat \rho = \bar \rho +  \epsilon \rho, \quad \hat T = \bar T +  \epsilon T,
 \quad u \sim (\epsilon g d)^{1/2}, \quad t \sim \left( \frac{\epsilon g}{d} \right)^{-1/2}, \quad \hat s = \bar s +  \epsilon s.
\label{eq:2_7}
\end{equation}
where $\bar p$, $\bar \rho$ and $\bar T$ comprise the adiabatic reference state.  Here $\bar s$ is simply a constant, and since 
$s =c_v \ln {\hat p}/{\hat \rho}^{\gamma}$ + const., and $\bar p /  {\bar \rho}^\gamma$ is constant, we obtain 
 (\ref{eq:2_4}b) by choosing the constant appropriately. 

\subsection{The adiabatic reference state}
 
The reference state is the adiabatic static equilibrium governed by $\mathrm{d}\bar{p}/\mathrm{d}z=-g\bar{\rho}$, 
$\bar{p}= R \bar{\rho} \bar{T}$ and $\bar p = K {\bar \rho}^\gamma$,
 where $R$ is the gas constant, $z=0$ is the bottom of the layer and $z=d$ the top. It follows that 
%
%\begin{subequations}
\begin{equation*}
\refstepcounter{equation}
\bar{T}=T_{B}\left(1-\theta\frac{z}{d}\right),\quad\bar{\rho}=\rho_{B}\left(1-\theta\frac{z}{d}\right)^{m}, \quad
\bar{p}=\frac{gd\rho_{B}}{\theta\left(m+1\right)}\left(1-\theta\frac{z}{d}\right)^{m+1}, 
\eqno{(\theequation \textit{a,b,c})}  \label{eq:2_8}
\end{equation*}
\begin{equation*}
\frac{gd}{c_p}=\Delta {\bar T}=T_{B}-T_{T}>0, \quad \theta=\frac{\Delta {\bar T}}{T_{B}}, \quad m=\frac{1}{\gamma -1},
\quad \Gamma= \frac{T_B}{T_T} = \frac{1}{1-\theta}, 
\eqno{(\theequation \textit{d,e,f,g})}
\end{equation*}
%%\end{subequations}
which defines $\theta$, and the polytropic index $m$. We use subscripts $T$ and $B$ to
denote conditions at the top and bottom boundary respectively. The temperature ratio $\Gamma>1$, is a convenient measure of 
the compressibility. $\Gamma \to 1$ is the Boussinesq limit, and highly compressible
layers have $\Gamma$ large. Note that $\Gamma^m = \rho_{B} / \rho_{T}$ is the ratio of the highest to lowest density
in the layer. The density ratio can be very large in astrophysical applications, the density of the bottom of the
solar convection zone being $\sim 10^6$ times the density at the top. Sometimes the number of density scale heights, $N_{\rho}$,
(the scale height being defined at the top of the layer) 
 that fit into
the layer is used to measure compressibility, $N_{\rho} = m(\Gamma -1)$.   

\subsection{The conduction state}

The adiabatic reference state satisfies $\nabla^2 {\bar T} =0$, but since it is isentropic is does not satisfy the entropy
boundary conditions. The anelastic conduction state is also a polytrope, but with a slightly more negative  temperature gradient, 
so ${\tilde T}_B=T_B, \  {\tilde T}_T < T_T$. The conduction state is 
\begin{equation*}
\refstepcounter{equation}
\tilde{T}=T_{B}\left(1-{\tilde \theta} \frac{z}{d}\right),\quad\tilde{\rho}=\rho_{B}\left(1-{\tilde \theta}
\frac{z}{d}\right)^{\tilde m}, \quad
\tilde {p}=\frac{gd\rho_{B}}{{\tilde \theta}\left({\tilde m}+1\right)}\left(1-{\tilde \theta}\frac{z}{d}\right)^{{\tilde m}+1}, 
\eqno{(\theequation \textit{a,b,c})}  \label{eq:2_9}
\end{equation*}
\begin{equation*}
{\widetilde {\Delta T}}=T_{B}-{\tilde T}_{T}>0, \quad {\tilde  \theta}=\frac{{\widetilde {\Delta T}}}{T_{B}}, 
\quad {\tilde m}=\frac{gd}{R {\widetilde {\Delta T}}} -1.
\eqno{(\theequation \textit{d,e,f})}
\end{equation*}
The small anelastic parameter $\epsilon$ is now defined as
\begin{equation}
\epsilon =  {\tilde \theta} \frac{{\tilde m}+1-\gamma {\tilde m}}{\gamma} 
 = -\frac{d}{T_{B}}\left[\frac{\mathrm{d}\tilde{T}}{\mathrm{d}z}+\frac{g}{c_{p}}\right]  \ll 1  ,
\label{eq:2_10}
\end{equation}
and the entropy in the conduction state is 
\begin{equation*}
\refstepcounter{equation}
{\tilde s}=c_{v} \ln\frac{\tilde{p}}{\tilde{\rho}^{\gamma}} + \mathrm{const} 
=\frac{\epsilon c_p}{\theta}
\ln\left(1-\theta\frac{z}{d}\right)+\mathrm{const}, \ \textrm{so} \ s = \frac{c_p}{\theta}
\ln\left(1-\theta\frac{z}{d}\right)+\mathrm{const}
\eqno{(\theequation )}  \label{eq:2_11}
\end{equation*}
which is the scaled entropy, see (\ref{eq:2_7}), correct to $O(\epsilon)$ since $\tilde \theta$ and $\theta$ differ by only $O(\epsilon)$. Since the boundaries have fixed entropy, the entropy at the boundaries in
the conduction state defines the entropy drop across the layer for all Rayleigh numbers, so
\begin{equation}
\Delta S =   \frac{ c_p}{\theta} \ln \Gamma = \frac{c_p \Gamma \ln \Gamma}{\Gamma-1}, 
 \label{eq:2_12}
\end{equation}
 relating to the entropy boundary conditions (\ref{eq:2_6}). Note that as our entropy variable $s$ is scaled by $\epsilon$,  the entropy drop is $O(\epsilon)$. 
Some anelastic papers take the conduction state as the reference state, and some take the adiabatic state as the 
reference state. Taking the conduction state as the reference state is appropriate when convection
near critical Rayleigh number is considered, but at the large Rayleigh numbers considered here, the conduction
state is less relevant.  Although the conduction state tends to the adiabatic reference state as $\epsilon \to 0$, the
 thermodynamic variables are not the same in the two systems: $T=0$ with respect  to the adiabatic state corresponds
to $T= T_B z /d$ if the conduction state is chosen as the reference state.

In equation (\ref{eq:2_1}) we have made use of (\ref{eq:2_4}b), (\ref{eq:2_8})  and (\ref{eq:2_10}) to write (Braginsky \& Roberts,
1995; Lantz \& Fan, 1999)
\begin{eqnarray}
-\frac{\nabla p}{\bar{\rho}}-\frac{\rho}{\bar{\rho}}g\hat{\mathbf{e}}_{z}  =  
-\nabla\left(\frac{p}{\bar{\rho}}\right)+\frac{g}{c_{p}}s\hat{\mathbf{e}}_{z} + O(\epsilon)  .\label{eq:2_13}
\end{eqnarray}

\subsection{The Nusselt and Rayleigh numbers in anelastic convection}

Next we consider the superadiabatic heat flux. The horizontal average at level $z$ is denoted by $\langle \ \rangle_{h}$. At the boundaries, all the heat is carried by conduction, and if
the total temperature $\hat T = \bar{T} + \epsilon T$, then the total heat flux at the boundaries is
$ -k d{\left\langle \hat T \right\rangle_h} /dz = -k d \bar{T} /dz - k  \epsilon d\left\langle T \right\rangle_h /dz$, but the superadiabatic part is obtained by subtracting off
the heat flux carried along the adiabat, so we let
\begin{equation}
\epsilon F^{super} = - \epsilon k \frac{d \left\langle T \right\rangle_h}{dz}\Big|_{z=0}, \quad {\rm so} \quad
 F^{super} =  -  k \frac{d \left\langle T \right\rangle_h }{dz}\Big|_{z=0}. 
\label{eq:2_14}
\end{equation}
The Nusselt number in anelastic convection is defined as the ratio of the superadiabatic heat
flux divided by the heat conducted down the conduction state superadiabatic gradient. 
Note that the flux conducted down 
the adiabatic gradient is ignored in this definition, so
\begin{equation}
Nu = \frac{F^{super} d}{k T_B}, \label{eq:2_15} 
\end{equation}
so $Nu$ is close to unity near onset, and is large in fully developed convection. 
For fixed entropy boundary conditions, the Rayleigh number is defined as 
\begin{equation}
Ra=\frac{g\Delta S d^{3}\rho_{B}^{2}}{\mu k}\approx\frac{c_{p}\Delta S \Delta {\bar T} d^{2}\rho_{B}^{2}}{\mu k}.\label{eq:2_16}
\end{equation}
The anelastic approximation is asymptotically valid in the limit $\epsilon \to 0$. Note that small superadiabatic temperature
gradient does not imply small Rayleigh number $Ra$, since the diffusion coefficients can be small, 
in fact to get $Ra \sim O(1)$ in the limit $\epsilon \to 0$ we must have
\begin{equation}
\frac{k}{\rho_B c_p} \sim \left( g d^3 \epsilon \right)^{1/2},  \quad  \frac{\mu}{\rho_B } \sim \left( g d^3 \epsilon \right)^{1/2}. \label{eq:2_17}
\end{equation}
allowing large but finite $Ra$ even when the superadiabatic gradient is small. 
To derive the anelastic equations (\ref{eq:2_1}) to (\ref{eq:2_5}), we insert (\ref{eq:2_7}) into the full compressible
equations and divide the momentum equation by $\epsilon$, the mass conservation equation by $\epsilon^{1/2}$ and the
energy equation by $\epsilon^{3/2}$. Having taken this limit, the parameter
$\epsilon$ no longer appears in our analysis. However, if anelastic work is compared to fully compressible situations,
then a finite value of $\epsilon$ must be chosen, and the anelastic results are only approximate, though 
there is a growing body of evidence that the anelastic approximation does capture the main features of subsonic compressible convection.

\subsection{High Rayleigh number convection}

\begin{figure}
\vspace{5mm}
\begin{centering}
\includegraphics[scale=0.18]{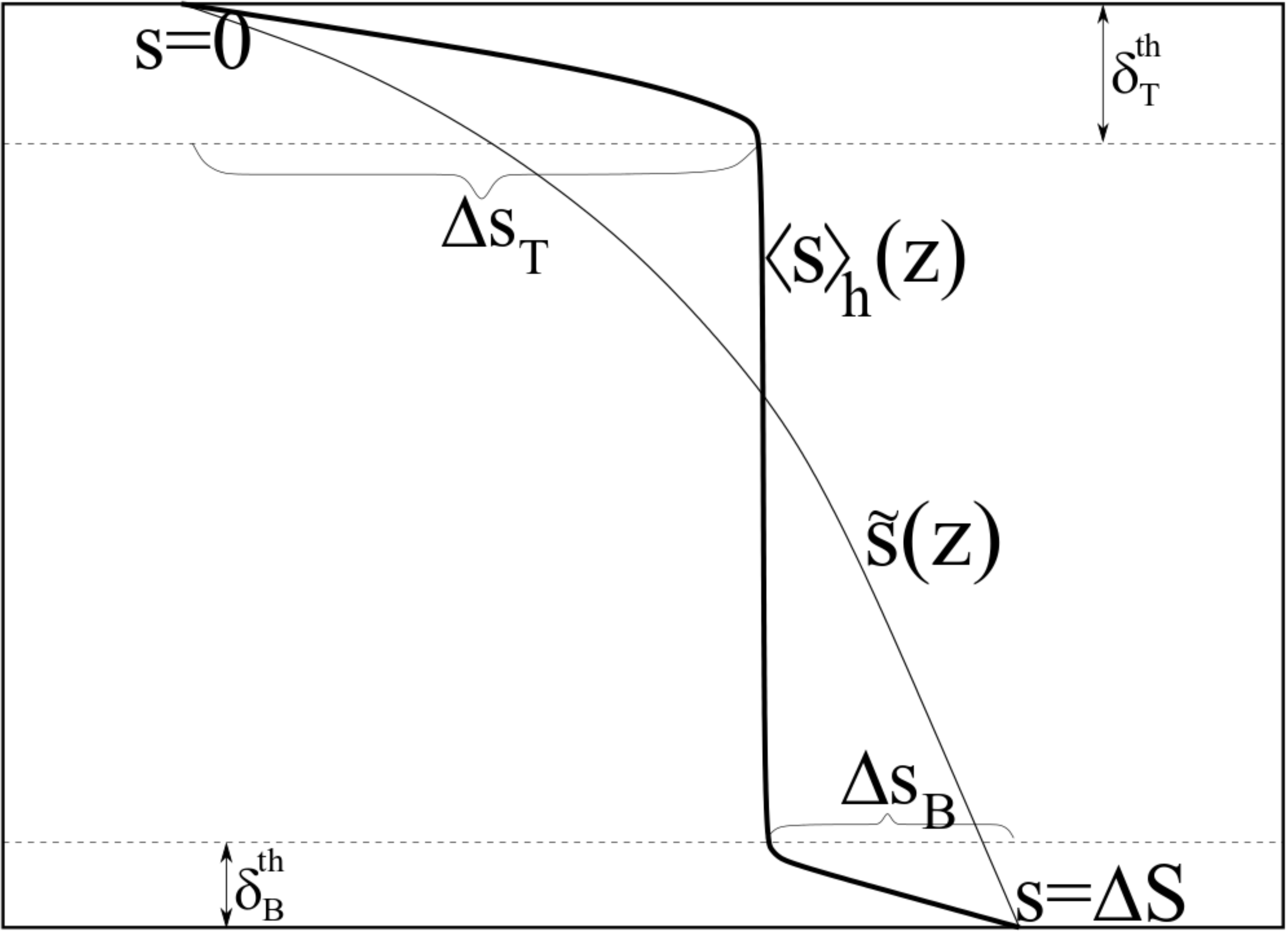} \hspace{5mm}  \includegraphics[scale=0.18]{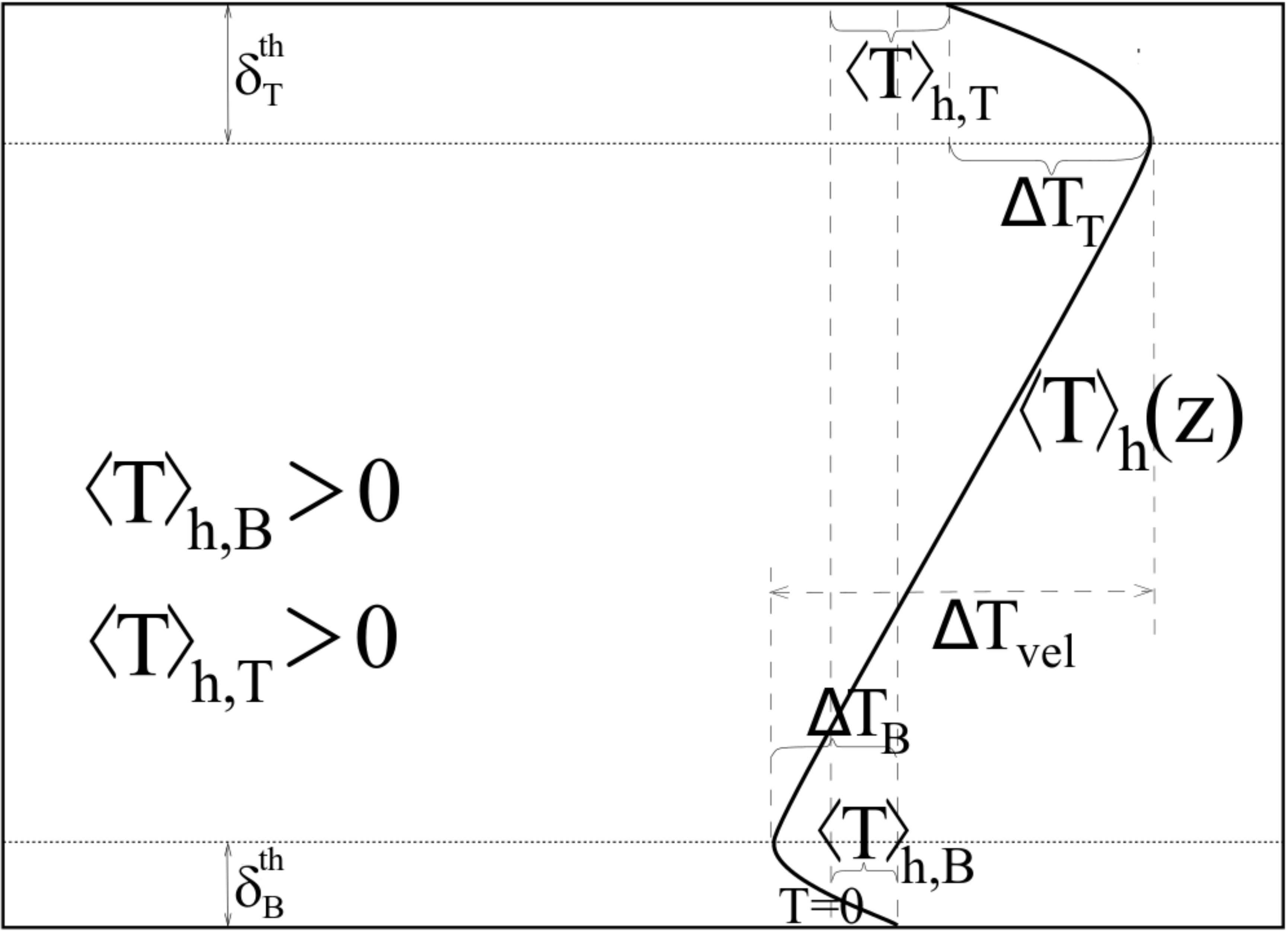} 
\par\end{centering}
\vspace{-43mm}
\hspace{-2mm} (a) \hspace{60mm} (b)

\vspace{15mm}
\hspace{-4mm} z \hspace{67mm} z

\vspace{21mm}
\hspace{30mm} $\langle s \rangle_h$ \hspace{55mm} $\langle T \rangle_h$  

\caption{(a) A schematic picture of the  entropy profile in developed convection.
(b) a schematic picture of the anelastic temperature perturbation in developed convection.}
\label{fig1}
\end{figure}

We have the following physical picture in mind. In strongly turbulent convection we
expect the entropy $s$ to be well-mixed away from boundary layers near $z=0,d$. 
We denote the global spatial average over the convecting layer by $\vert \vert \ \ \vert \vert$ and
the horizontal average at level $z$ by $\langle \ \rangle_h$.
The total entropy drop is the conduction state value $\Delta S = c_p \ln \Gamma / \theta.$
Since the entropy is constant in the bulk interior, we define the entropy drops $\Delta S_B$
and $\Delta S_T$ across the bottom and top boundary layers respectively. These will not be equal,
with $\Delta S_T$ normally considerably larger than $\Delta S_B$. We must however have
\begin{equation}
\Delta S_B + \Delta S_T = \Delta S.
\label{eq:2_18}
\end{equation}
We consider only the case where both the top and bottom boundary layers are laminar. At extremely
high $Ra$ these layers may become turbulent as can happen in the Boussinesq case. The laminar
boundary layer case is simpler, and gives predictions which can be broadly compared with numerical simulations,
though it is difficult for numerical simulations to get into the fully developed large Rayleigh and Nusselt number
regime we are aiming at here.
A schematic picture of the horizontally-averaged entropy profile expected in
highly supercritical anelastic convection is shown in figure \ref{fig1}(a).

Since the heat flux through the boundary layers is determined by thermal diffusion rather
than entropy diffusion, we need to express the temperature jumps across
the thermal boundary layers in terms of the entropy jumps. From (\ref{eq:2_4})
we obtain
\begin{equation*}
\refstepcounter{equation}
\frac{\left(\Delta\rho\right)_{i}}{\rho_{i}}\approx\frac{1}{\gamma-1}\left[\frac{\left(\Delta T\right)_{i}}{T_{i}}-\gamma\frac{\left(\Delta s\right)_{i}}{c_{p}}\right],\qquad\frac{\left(\Delta p\right)_{i}}{p_{i}}\approx\frac{\gamma}{\gamma-1}\left[\frac{\left(\Delta T\right)_{i}}{T_{i}}-\frac{\left(\Delta s\right)_{i}}{c_{p}}\right],
\eqno{(\theequation \textit{a,b})} \label{eq:2_19} 
\end{equation*}
where the  $\Delta$ quantities refer to the jump in the horizontally averaged value across the boundary layer 
and the subscript $i$ stands either for $B$ or $T$. We also define the thermal and viscous 
boundary layer thicknesses, $\delta_i^{th}$ and $\delta_i^{\nu}$ with $i=B,T$,
which we use to obtain scaling estimates. Numerical simulations indicate that the horizontal velocity 
\begin{equation}
U_H = \left( \left\langle u_x^2 \right \rangle_h + \left\langle u_y^2 \right \rangle_h \right)^{1/2}
\label{eq:2_20}
\end{equation}
has local maxima close to both boundaries (see e.g. figure  \ref{fig3}(b) below), so these maxima are a convenient way to define the velocity jumps across the viscous boundary layers, $\Delta U_i$, so
\begin{equation}
\Delta U_B = U_H(z=z_{max,B}), \quad \Delta U_T = U_H(z=z_{max,T}),
\label{eq:2_21}
\end{equation}
where $z=z_{max,B}$, $ z=z_{max,T}$ are the locations of the local maxima. The thermal boundary
layer thickness for the entropy, $\delta_i^{th}$, and the viscous boundary layer thickness, $\delta_i^{\nu}$, can be defined as
\begin{equation}
\delta_i^{th} = \left\{  - \frac{1}{\Delta S_i}   \frac{d \left\langle S \right\rangle_h}{dz} {\Big \vert}_{z=z_i} \right\}^{-1}, \quad \delta_i^{\nu} = \left\{  \pm \frac{1}{\Delta U_i}   \frac{d U_H }{dz} {\Big \vert}_{z=z_i} \right\}^{-1}, \quad z_i = z_B, \, z_T
\label{eq:2_22}
\end{equation}
%

% FIGURE 1
In the boudary layers, the dominant balance in the $z$-component of the Navier-Stokes equation occurs
between the pressure gradient and the buoyancy force. Mass conservation in the boundary layers means 
$u_{z,i} \sim O(\delta_i^{\nu})$ so the vertical component of inertia is small. The boundary layers are therefore 
approximately hydrostatic,
\begin{equation}
\left(\Delta p\right)_{i}\approx\frac{g}{c_{p}}\rho_{i}\Delta s_{i}\delta_{i}^{th}.\label{eq:2_23}
\end{equation}
Inserting (\ref{eq:2_23}) into (\ref{eq:2_19}b)
leads to
\begin{equation}
\frac{\left(\Delta T\right)_{i}}{T_{i}}\approx\frac{\left(\Delta s\right)_{i}}{c_{p}}\left(1+\theta\frac{\delta_{i}^{th}}{d}\frac{T_{B}}{T_{i}}\right).\label{eq:2_24}
\end{equation}
Typically the term $(\theta\delta_{i}^{th}T_{B})/(T_{i}d)$ resulting
from the pressure jump across the boundary layers is expected to be
small because the boundary layer is thin. However, in simulations where the Rayleigh
number is bounded above by numerical constraints, the top boundary layer may
not be as thin as desired for accurate asymptotics to apply, and the factor $T_B/T_T$
can be large in layers containing many scale heights. 
We refer to the case  where the pressure term is not negligible as the compressible boundary layer case. However, in this work  we assume the  boundary layers are incompressible, which is valid
provided  $T_B/T_T$ remains 
finite as the Rayleigh number increases and the boundary layers become very thin. Then the pressure
fluctuation term is constant in both boundary layers, so that in the boundary layers
\begin{equation}
\frac{\left(\Delta T\right)_{i}}{T_{i}}\approx\frac{\left(\Delta s\right)_{i}}{c_{p}}, \ \
\textrm{and} \ \  \frac{\left(\nabla T\right)_{i}}{T_{i}}\approx\frac{\left(\nabla s \right)_{i}}{c_{p}}
\label{eq:2_25}
\end{equation}
 and defining the temperature boundary layer thicknesses similarly to (\ref{eq:2_22}),
\begin{equation}
\delta_i^{T} = \left\{  - \frac{1}{\Delta T_i}   \frac{d \left\langle T \right\rangle_h}{dz} {\Big \vert}_{z=z_i} \right\}^{-1}, 
\label{eq:2_26}
\end{equation}
the temperature boundary layer thicknesses are the same as the entropy boundary layer thicknesses. Note that in the compressible boundary layer case the entropy and temperature boundary layer thicknesses will be different. For incompressible boundary layers and high Rayleigh number, the Nusselt number can be written in terms of the boundary layer thicknesses, using (\ref{eq:2_15}), (\ref{eq:2_14}), (\ref{eq:2_26}) and (\ref{eq:2_25}),
\begin{equation}
Nu =   \frac{d}{\delta^{th}_T} \frac{\Delta S_T}{\Gamma c_p}  =  \frac{d}{\delta^{th}_B} \frac{\Delta S_B}{c_p} .
\label{eq:2_27}
\end{equation}
 
In figure \ref{fig1}(b) we sketch the horizontally-averaged anelastic temperature perturbation $\langle T \rangle_h$.
This is sometimes referred to as the superadiabatic temperature (e.g. Verhoeven \textit{et al.} 2015). 
Note that with our constant entropy boundary conditions, $\langle T \rangle_h$ is not
zero at the boundaries. We show in appendix A that the offsets, $\langle T \rangle_{hB}$ at $z=0$ and   $\langle T \rangle_{hT}$ at $z=d$,
are both positive and we show also that in the bulk, turbulent pressure effects 
make the gradient of $T$ positive as shown in figure  \ref{fig1}(b). This means
that the total horizontally averaged temperature gradient in the presence of convection
is less negative than the adiabatic
reference state, so that on horizontal average the layer is subadiabatically stratified (e.g. Korre \textit{et al.} 2017).
Of course, in some parts of the layer the local temperature gradient must be superadiabatic 
to drive the convection, but other parts are subadiabatic so that the horizontal 
average can be subadiabatic. 

To obtain the anelastic temperature fluctuation as sketched in figure  \ref{fig1}(b), we need to make use of 
equations (\ref{eq:2_4}a) and (\ref{eq:2_4}b), so we need to make a specific choice for entropy at the boundaries. 
Here we have chosen to take the entropy at the top boundary as zero, so the entropy at
the bottom boundary is $s=\Delta S$. A different choice of entropy constant adds an
easily found function of $z$ to $T$, $\rho$ and $p$ but this does not affect the velocity field obtained.
One further point is that if (\ref{eq:2_1}) is horizontally averaged, the horizontal average satisfies
a first order differential equation in $z$ (see appendix A for details), so a boundary
condition on $ \langle p \rangle_h$ is required. Here we choose the natural condition that the anelastic density
perturbation vanishes when integrated over the layer, that is 
\begin{equation}
 \vert \vert \ \rho \ \vert \vert = 0 \quad \Rightarrow \quad \langle p \rangle_{h,T} = \langle p \rangle_{h,B}. 
\label{eq:2_28}
\end{equation}      
This means that the total mass in the layer is the same as in the adiabatic reference state.
As we see in appendix A, this means the horizontally averaged anelastic pressure perturbations
at the top and bottom of the layer must be equal.

%%%%%%%%%%%%%%%%%%%%%%%%%%%%%%%%%%%%
%  3. Energy and entropy production integrals
%%%%%%%%%%%%%%%%%%%%%%%%%%%%%%%%%%%%%%%%%%

\section{\label{section3}Energy and entropy production integrals}

Understanding the energy transfer and entropy production in convective flow
is the key to understanding the physics of compressible convection.
Therefore we derive now a few exact relations which will allow us
to study some general aspects of the dynamics of developed compressible
convection.  
We assume any initial transients in
the convection have been eliminated, and we are in a statistically steady state. Formally,
this means we consider time-averaged quantities throughout the paper.

\subsection{Energy balance}

By multiplying the Navier-Stokes equation (\ref{eq:2_1}) by $\bar{\rho} {u}$ and averaging over the entire 
volume (recalling that horizontal averages of $x$ and $y$ derivatives are zero) we obtain the following relation
\begin{equation}
\frac{g}{c_{p}} \vert \vert \bar{\rho}u_{z}s \vert \vert =\mu\vert \vert q\vert \vert ,\label{eq:3_1}
\end{equation}
stating that the total work per unit volume of the buoyancy
force is equal to the total
viscous dissipation in the fluid per unit volume. In deriving (\ref{eq:3_1}) use has been made of the no-slip
boundary conditions and equation (\ref{eq:2_2}).

We derive the superadiabatic heat flux in the system at every $z$
by averaging over a horizontal plane and integrating the heat equation (\ref{eq:2_3}) from $0$ to
$z$,
\begin{eqnarray}
F^{super}&=& -  k \frac{d\left\langle T \right\rangle_h}{dz}\Big|_{z=0}  
=\left\langle \bar{\rho}\bar{T}u_{z}s\right\rangle _{h}
-  k \frac{d\left\langle T \right\rangle_h }{dz}\nonumber \\
&+&\frac{g}{c_p}\int_{0}^{z}\left\langle \bar{\rho}u_{z}s\right\rangle _{h}\mathrm{d}z-\mu\int_{0}^{z}\left\langle q\right\rangle_h \mathrm{d}z
  -2\mu\left[ \left\langle  u_{z} \frac{du_{z}}{\mathrm{d}z}  \right\rangle_{h}-\frac{m\Delta {\bar T}}{\bar{T} d}\left\langle u_{z}^{2}\right\rangle _{h}\right]. \label{eq:3_2}
\end{eqnarray}
In deriving this expression, we have made use of (\ref{eq:2_8}a,d) and
\begin{equation}
(\partial_j u_i)(\partial_i u_j)- (\nabla \cdot {\bf u})^2 = \partial_j \partial_i (u_i u_j) - 2 \partial_j ( u_j  \nabla \cdot {\bf u})
= \partial_j \partial_i (u_i u_j) + 2 \partial_j ( u_j  u_z \frac{m \Delta {\bar T}}{\bar T}) 
\label{eq:3_3}
\end{equation}
since the continuity equation gives 
\begin{equation}
\nabla \cdot {\bf u} = \partial_i u_i  = \frac{ u_z m \Delta {\bar T}}{ {\bar T} d},
\label{eq:3_4}
\end{equation}
 and we recall that  $x$ or $y$ derivatives vanish on horizontal averaging.
As $z \to d$ all terms with a factor $u_z$ tend to zero,  so on using (\ref{eq:3_1})  we obtain the overall energy balance equation,
\begin{equation}
F^{super}= -  k \frac{d\left\langle T \right\rangle_h}{dz}\Big|_{z=0}=  -  k \frac{d\left\langle T \right\rangle_h}{dz}\Big|_{z=d},\label{eq:3_5}
\end{equation}
thus in a stationary state the heat flux entering the fluid
volume at $z=0$ must be equal to the outgoing heat flux $z=d$.

\subsection{Entropy balance}

This energy balance equation alone is not very helpful for
evaluating the Nusselt number. We need the entropy balance equation,
obtained by dividing the energy equation (\ref{eq:2_3}) by $\bar{T}$, horizontally averaging,  and integrating
from $0$ to $z$,
\begin{eqnarray}
 \left\langle \bar{\rho}u_{z}s\right\rangle_h &=& - \frac{k}{T_B} \frac{\mathrm{d}}{\mathrm{d}z} \left\langle T \right\rangle_h \Big|_{z=0}
+ \frac{k}{\bar T} \frac{\mathrm{d}}{\mathrm{d}z} \left\langle T \right\rangle_h \Big|_z  
- \int_0^z \frac{k \Delta {\bar T}}{{\bar T}^2 d} \frac{\mathrm{d}}{\mathrm{d}z} \left\langle T \right\rangle_h \, dz 
+ \int_0^z \frac{\mu}{\bar T} \left\langle q \right\rangle_h \, dz  \nonumber \\
 &+& \int_0^z \frac{\mu}{\bar T} \left\langle \partial_j(\partial_i (u_i u_j)) -2\partial_j(u_j(\partial_iu_i)) \right\rangle_h \, dz
\label{eq:3_6}
\end{eqnarray}
\noindent where use has been made of equation (\ref{eq:3_3}). The overall entropy balance equation
comes from taking the limit $z \to d$ of (\ref{eq:3_6}), noting $u_z \to 0$ in this limit,
\begin{eqnarray}
  \frac{k}{T_B} \frac{\mathrm{d}}{\mathrm{d}z} \left\langle T \right\rangle_h \Big|_{z=0}
 - \frac{k}{T_T} \frac{\mathrm{d}}{\mathrm{d}z} \left\langle T \right\rangle_h \Big|_{z=d}  &=& 
- \int_0^d \frac{k \Delta {\bar T}}{{\bar T}^2 d} \frac{\mathrm{d}}{\mathrm{d}z} \left\langle T \right\rangle_h \, dz 
+ \int_0^d \frac{\mu}{\bar T} \left\langle q \right\rangle_h \, dz  \nonumber \\
\label{eq:3_7}
\end{eqnarray}
Equations  (\ref{eq:3_6}, \ref{eq:3_7}) look complicated, but they simplify considerably
when the top and bottom boundary layers are thin. We start with   (\ref{eq:3_6}), which we write as
\begin{eqnarray}
 S_{conv} = S_{diff} + S_{visc} 
\label{eq:3_8}
\end{eqnarray}   
where 
\begin{eqnarray}
 S_{conv} =  \left\langle \bar{\rho}u_{z}s\right\rangle_h, 
\label{eq:3_9}
\end{eqnarray}
the net entropy carried out of the region $(0,z)$ by the convective velocity at level $z$,
\begin{eqnarray}
  S_{diff} =- \frac{k}{T_B} \frac{\mathrm{d}}{\mathrm{d}z} \left\langle T \right\rangle_h \Big|_{z=0}
+ \frac{k}{\bar T} \frac{\mathrm{d}}{\mathrm{d}z} \left\langle T \right\rangle_h \Big|_z  
- \int_0^z \frac{k \Delta {\bar T}}{{\bar T}^2 d} \frac{\mathrm{d}}{\mathrm{d}z} \left\langle T \right\rangle_h \, dz
\label{eq:3_10}
\end{eqnarray}   
so that $S_{diff}$ is the entropy balance in region $(0,z)$ of the entropy change due to thermal diffusion. 
This is divided into the first term,
which represents the positive entropy being conducted into our region at the bottom boundary (the gradient of $\langle T \rangle_h$
is negative there, see figure 1b), the second term is the entropy conducted across level $z$, and the third term is
the entropy production by internal diffusion in our given region. By looking at figure 1b it is apparent that
when the  boundary layers are thin, the first term is much larger than the other two except when $z$ is in the
boundary layers. If $z$ is in the bulk, the gradient of $\langle T \rangle_h$ is $O(\Delta {\bar T} / d)$ whereas at the boundary it is
 $O(\Delta {\bar T} / \delta^{th})$, much larger since the boundary layer is thin. 
The third term is always small compared to the first, because the gradient is $O(\Delta {\bar T} / d)$ outside the boundary layers. The integrand is of order 
$O(\Delta {\bar T} / \delta^{th})$ in the boundary layers, but because they are thin this only contributes a small amount
to the integral. So when the boundary layers are thin
\begin{eqnarray}
  S_{diff}  \approx - \frac{k}{T_B} \frac{\mathrm{d}}{\mathrm{d}z} \left\langle T \right\rangle_h \Big|_{z=0}
\quad \textrm{if $z$ is in the bulk},
\label{eq:3_11}
\end{eqnarray} 
\begin{eqnarray}
S_{diff}  \approx - \frac{k}{T_B} \frac{\mathrm{d}}{\mathrm{d}z} \left\langle T \right\rangle_h \Big|_{z=0}
+ \frac{k}{T_T} \frac{\mathrm{d}}{\mathrm{d}z} \left\langle T \right\rangle_h \Big|_{z=d} \quad \textrm{if $z=d$ . }
\label{eq:3_12}
\end{eqnarray}  
We now turn to 
\begin{equation}
S_{visc} =
 \int_0^z \frac{\mu}{\bar T} \left\langle q \right\rangle_h \, dz  
 + \int_0^z \frac{\mu}{\bar T} \left\langle \partial_j(\partial_i (u_i u_j)) -2\partial_j(u_j(\partial_iu_i)) \right\rangle_h \, dz .
\label{eq:3_13}
\end{equation}

Because of the horizontal averaging, and using equations (\ref{eq:2_2}), (\ref{eq:2_8}) and (\ref{eq:3_4}), the second integral can be written
\begin{eqnarray}
\int_0^z 
\frac{\mu}{\bar T} \left\langle \partial_j(\partial_i (u_i u_j)) - 2\partial_j(u_j(\partial_iu_i)) \right\rangle_h \, dz = 
\qquad \qquad \nonumber \\
 \int_{0}^{z} \frac{\mu}{\bar T} \left[ \frac{\partial^2}{\partial z^2}  \left\langle u_{z}^{2}\right\rangle_h
- \frac{m \Delta {\bar T}}{{\bar T} d} \frac{\partial}{\partial z} \left\langle u_{z}^{2}\right\rangle_h
-  \frac{m (\Delta {\bar T})^2}{{\bar T}^2 d^2} \left\langle u_{z}^{2}\right\rangle_h  \right] \, dz .
 \label{eq:3_14}
\end{eqnarray}  
We now consider the magnitude of the terms in equation (\ref{eq:3_13}). If the 
root mean square velocity in the bulk is $U$, we expect the horizontal velocity 
to vary from 0 to $O(U)$ across the boundary layers of thickness $\delta_i^{\nu}$, so the velocity
gradients appearing in $q$ are of magnitude $O(U/\delta_i^{\nu})$. $q$ itself is therefore
$O(U^2/{(\delta_i^{\nu}})^2)$, and since the boundary layers are thin their contribution to
the first integral in $S_{visc}$ is $O(\mu U^2/ {\bar T} \delta_i^{\nu})$. In the boundary layers 
$u_z$ is small, but in the bulk we expect $u_z$ to be $O(U)$ and so $\langle u_z^2 \rangle_h$ is $O(U^2)$. 
Because $\langle u_z^2 \rangle_h$  is horizontally averaged, it will vary on a length-scale $d$ with $z$, so the
gradient of $ \langle u_z^2 \rangle_h$ is $O(U^2/d)$ and the second derivative is  $O(U^2/d^2)$.  From (\ref{eq:3_14}), the order of magnitude of
the second term in  (\ref{eq:3_13}) is then $O(\mu U^2/ T d)$, which is $O( \delta_i^{\nu}/d)$ smaller
than the contribution from the term due to $q$. Therefore provided the Rayleigh number is high enough for the boundary layers
to be thin, equation  (\ref{eq:3_6}) is asymptotically equivalent to
\begin{eqnarray}
   \left\langle \bar{\rho}u_{z}s\right\rangle_h \sim  -  \frac{k}{T_B} \frac{\mathrm{d}}{\mathrm{d}z} \left\langle T \right\rangle_h \Big|_{z=0}
  +  \int_0^z \frac{\mu}{\bar T} \left\langle q \right\rangle_h \, dz .
\label{eq:3_15}
\end{eqnarray}
when $z$ is in the bulk.
Note that this simplification still holds if the dissipation in the bulk is larger than the dissipation in the boundary layers,
which can happen, as noted by Grossmann \& Lohse (2000). When the dissipation is primarily in the boundary layers,
the left-hand-side of  ( \ref{eq:3_15}) is constant in the bulk, which we exploit later. 
In either case, as $z \to d$ we get 
\begin{eqnarray}
  \frac{k}{T_B} \frac{\mathrm{d}}{\mathrm{d}z} \left\langle T \right\rangle_h \Big|_{z=0}
 - \frac{k}{T_T} \frac{\mathrm{d}}{\mathrm{d}z} \left\langle T \right\rangle_h \Big|_{z=d} 
= \frac{F^{super} \Delta {\bar T}}{T_B T_T}  \sim
   \int_0^d \frac{\mu}{\bar T} \left\langle q \right\rangle_h \, dz.  \nonumber \\
\label{eq:3_16}
\end{eqnarray}
Note also that in these expressions, the term $(\nabla \cdot {\bf u})^2/3$ in equation (\ref{eq:2_5}) makes a 
negligible contribution to (\ref{eq:3_16}) compared to the gradient terms by using  (\ref{eq:3_4}).

\subsection{The Boussinesq limit}
      
At first sight, it appears that our entropy balance formulation of the equation for dissipation (\ref{eq:3_7})  is fundamentally different from that used by Grossman \& Lohse (2000) in the Boussinesq case, who start from equation (2.5) of Siggia (1994), 
\begin{equation}
(Nu-1) Ra = \langle(\nabla v)^2 \rangle, \ \textrm{or} \quad 
g \alpha \kappa \Delta T (Nu-1) = \nu \int_0^d \langle q \rangle_h \, dz 
\label{eq:3_17}
\end{equation} 
in our dimensional units. Here $\Delta T$ is the superadiabatic temperature difference between the boundaries. In the Boussinesq limit $\Gamma \to 1$, the basic state temperature and density tend to constant values $T_B$ and $\rho_B$ respectively, so the thermal diffusivity $\kappa$ and kinematic viscosity $\nu$ are constants, $\kappa=k / \rho_B c_p$ and 
$\nu= \mu / \rho_B$. For a perfect gas the coefficient of expansion $\alpha=1/T_B$. In this subsection we show that (\ref{eq:3_17}) is in fact the Boussinesq limit of the entropy balance equation (\ref{eq:3_7}), which we use to derive the scaling laws in \S6 below. Our entropy balance formulation is a generalization of the Grossmann \& Lohse (2000) formulation, which is now seen to be a limiting case of the more general entropy balance approach.

Following Spiegel \& Veronis (1960) we note that from the $z$-component of (\ref{eq:2_1}) $p/{\bar \rho} d \sim g s /c_p$, so
 \begin{equation}
\frac{p}{\bar p} \sim \frac {g {\bar \rho} d}{{\bar p}} \frac{s}{c_p}  \sim \frac{d}{H} \frac{s}{c_p}
\label{eq:3_18}
\end{equation}
where $H$ is the pressure scale height. In the Boussinesq limit $d/H$ becomes small; the Boussinesq 
limit $\Gamma \to 1$ is the thin layer limit (Spiegel \& Veronis, 1960). Then (\ref{eq:2_4}a,b) become
 \begin{equation}
\frac{T}{T_B} \sim - \frac {\rho} { \rho_B} , \quad \frac{s}{c_p} \sim \alpha T,
\label{eq:3_19}
\end{equation}
so the entropy variable becomes equivalent to the superdiabatic temperature variable. The entropy jump $\Delta S$
across the layer can be written as a superadiabatic temperature jump $\Delta T = \Delta S/ \alpha c_p$.
 As $\Gamma \to 1$, from (\ref{eq:2_12}) $\Delta S / c_p \to 1$ so $\Delta T \to 1/ \alpha = T_B$. From energy conservation
(\ref{eq:3_5}) the gradient of $\langle T \rangle_h$ is the same at the top and bottom of the layer, so (\ref{eq:3_7})
can be written
 \begin{equation}
  -\frac{\Delta {\bar T}}{T_B^2} \frac{d}{dz} \langle T \rangle_h {\Big \vert}_{z=0} =
 - \frac{k \Delta {\bar T} }{T_B^2 d} \int_0^d \frac{d}{dz} \langle T \rangle_h \, dz +  \frac{\mu}{T_B}  \int_0^d \langle q \rangle_h \, dz,
\label{eq:3_20}
\end{equation} 
using the constancy of $\bar T$ in the Boussinesq limit.
From (\ref{eq:2_14}) and (\ref{eq:2_15}) 
\begin{equation}
  Nu = -\frac{d}{T_B} \frac{d}{dz} \langle T \rangle_h {\Big \vert}_{z=0} \ \textrm{or} \quad
  Nu = -\frac{d}{\Delta T} \frac{d}{dz} \langle T \rangle_h {\Big \vert}_{z=0}, 
\label{eq:3_21}
\end{equation} 
which is the familiar form of the Boussinesq Nusselt number, the ratio of the total heat flux at the bottom to the conducted heat flux $- k \Delta T / d$. From (\ref{eq:2_8}d) $\Delta {\bar T}$ can be written $g d / c_p$ so (\ref{eq:3_20}) becomes
\begin{equation}
  \frac{k}{c_p T_B} Nu g \alpha \Delta T =   \frac{k}{c_p T_B} g \alpha \Delta T + \frac{\mu}{T_B} 
\int_0^d \langle q \rangle_h \,dz  \  \  \textrm{or} \ \ 
  (Nu - 1) g a \Delta T \kappa = \nu \int_0^d \langle q \rangle_h \,dz , 
\label{eq:3_22}
\end{equation} 
which is (\ref{eq:3_17}), showing that the dissipation integral which plays a key role in Grossmann \& Lohse's (2000)
analysis is indeed the Boussinesq limit of the entropy balance equation (\ref{eq:3_7}). 

%%%%%%%%%%%%%%%%%%%%%%%%%%%%%%%%%%%%%%%%%%%%%%%%%%%%%
%  4. The boundary layers and Prandtl number effects
%%%%%%%%%%%%%%%%%%%%%%%%%%%%%%%%%%%%%%%%%%%%%%%%%%%%%%%

\section{\label{section4}The boundary layers and Prandtl number effects}

%FIGURE 2
\begin{figure}
\vspace{5mm}
\begin{centering}
\includegraphics[scale=0.2]{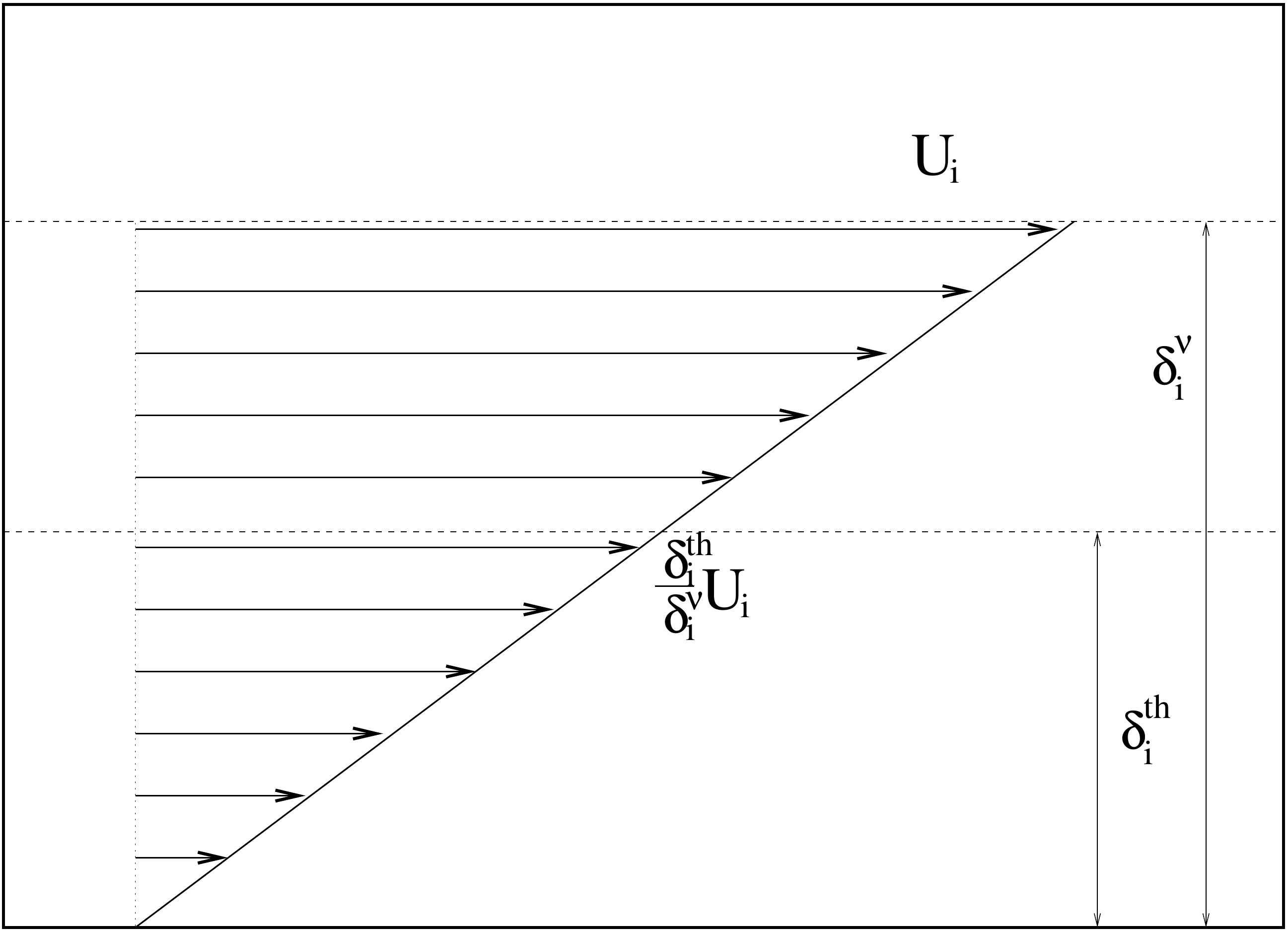} \hspace{5mm}  \includegraphics[scale=0.18]{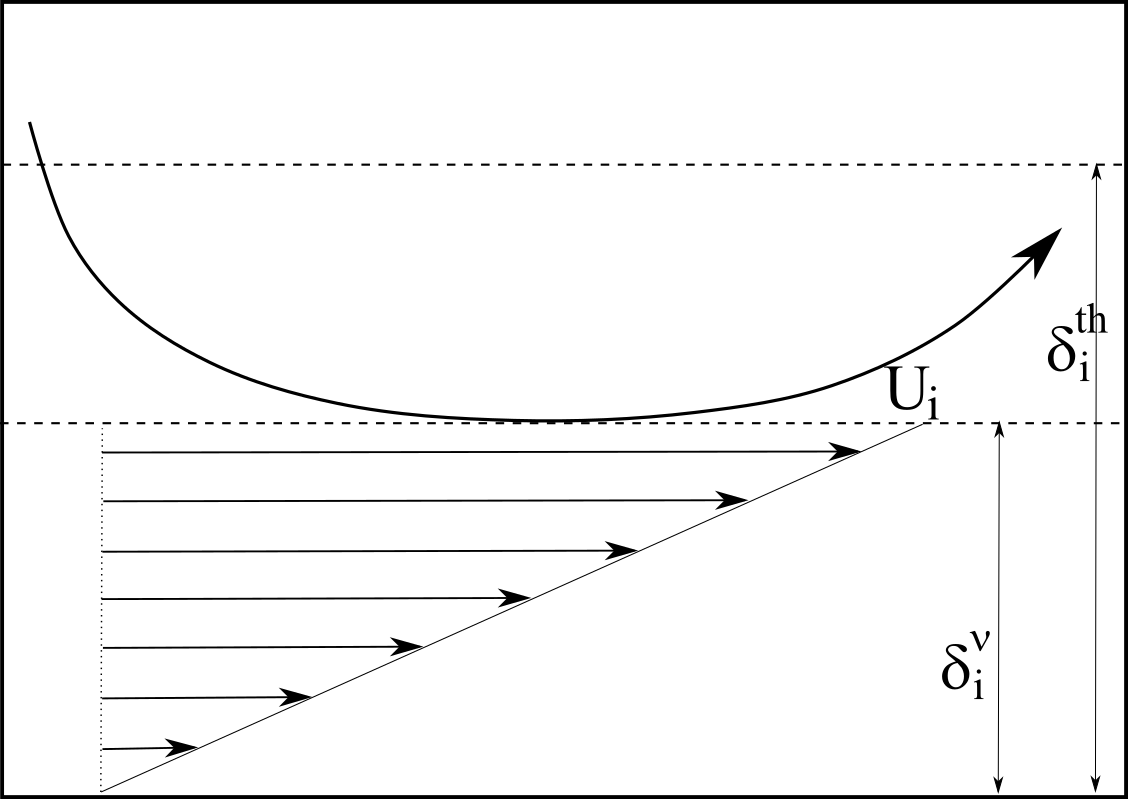}
\par\end{centering}

\vspace{-40mm}
\hspace{3mm} (a) \hspace{56mm} (b)

\vspace{37mm}

\caption{(a) Thermal and viscous boundary layers in the case $Pr >1$. The thermal diffusion is smaller, so the thermal
boundary layer is nested inside the viscous boundary layer. (b) The case $Pr < 1$, where the viscous boundary layer is nested inside 
the thermal boundary layer.  }
\label{fig2}
\end{figure}
As in the Boussinesq case, the thermal and viscous boundary layers can be nested inside each other when the Prandtl number is different from
unity. 

A central idea in the theory of fully developed Boussinesq
convection is based on the assumption that the structure of turbulent
convective flow is always characterized by the presence of a large-scale
convective roll called the \emph{wind of turbulence}, Grossmann \& Lohse (2000). This idea, which
in the non-stratified case stems from vast numerical and experimental
evidence, is retained in the case of anelastic convection. However,
the significant stratification in the anelastic case breaks the Boussinesq
up-down symmetry, and thus we must distinguish between the magnitude
of the wind of turbulence near the bottom of the bulk and its magnitude
near the top of the bulk, denoted by $U_{B}$ and $U_{T}$ respectively,
which can now significantly differ. So the label $U_{i}$ can denote either the
horizontal velocity just outside the top or bottom viscous boundary layers. We also assume that 
this wind of turbulence forms boundary layers with a horizontal length scale comparable to
the layer depth $d$. It is of course an assumption that such layers form in anelastic convection, but
they are observed to occur in incompressible flow, and the limited simulations we have available gives this idea some support. Whereas the results in \S2 and \S3 are asymptotically valid in the anelastic framework in the limit of large $Ra$, what follows is dependent on the Grossmann-Lohse (2000) approach being valid for anelastic convection. 

The Prandtl number is a constant in this problem, given by
\begin{equation}
Pr = \frac{\mu c_p}{k}.
\label{eq:4_1}
\end{equation}
In figure (\ref{fig2}a) the high Prandtl number case is shown, with the thinner thermal boundary layer nested inside the viscous boundary
layer. The velocity  at the edge of the thermal boundary layer is then $\delta_i^{th} U_i/ \delta_i^{\nu}$, assuming the velocity falls off linearly inside the viscous boundary layer over the
thinner thermal boundary layer. In the boundary layers, advection balances diffusion, so from
the momentum equation (\ref{eq:2_1}) we estimate that 
\begin{equation}
\frac{\rho_i U_i^2}{d} \sim \frac{\mu U_i }{\left(\delta_i^{\nu} \right)^2 }
\ \ \textrm{so} \ \ U_i \sim \frac{\mu d}{\rho_i \left(\delta_i^{\nu} \right)^2 }.
\label{eq:4_2}
\end{equation}
For the entropy boundary layers, from (\ref{eq:2_3})
\begin{equation}
\frac{\rho_i T_i U_i s}{d} \frac{\delta_i^{th}}{\delta_i^{\nu}}  \sim k \nabla^2  T 
\approx \frac{k T_i}{c_p} \nabla^2 s \sim \frac{k T_i s}{c_p  \left(\delta_i^{th} \right)^2},
\ \ \textrm{so} \ \ U_i = \frac{k d\delta_i^{\nu}}{\rho_i c_p \left(\delta_i^{th} \right)^3},
\label{eq:4_3}
\end{equation} 
where (\ref{eq:2_25}) has been used and the factor $\delta_i^{th}/{\delta_i^{\nu}}$ arises because the horizontal velocity is reduced because the entropy boundary layer is thinner than the viscous boundary layer. Dividing (\ref{eq:4_2}) by (\ref{eq:4_3}) we obtain
\begin{equation}
\frac{\delta_i^{\nu}}{\delta_i^{th}} \sim Pr^{1/3}
\label{eq:4_4}
\end{equation}
giving the ratio of the viscous to thermal boundary layer thickness for the high Prandtl number case.
Note that although the top and bottom boundary layers have different thicknesses, this ratio is the same for both layers.

For the low Prandtl number case, the viscous boundary layer lies inside the thermal boundary layer,
see figure (\ref{fig2}b). Now the velocity at the edge of the thermal boundary layer is the same as 
that at the edge of the viscous boundary layer, so the velocity reduction factor  $\delta_i^{th} / \delta_i^{\nu}$ is no longer required, so (\ref{eq:4_3}) becomes
\begin{equation}
\frac{\rho_i T_i U_i s}{d}  \sim k \nabla^2  T 
\approx \frac{k T_i}{c_p} \nabla^2 s \sim \frac{k T_i s}{c_p  \left(\delta_i^{th} \right)^2},
\ \ \textrm{so} \ \ U_i = \frac{k d }{\rho_i c_p \left(\delta_i^{th} \right)^2},
\label{eq:4_5}
\end{equation}
giving the ratio of the boundary layer thicknesses as
\begin{equation}
\frac{\delta_i^{\nu}}{\delta_i^{th}} \sim Pr^{1/2}
\label{eq:4_6}
\end{equation}
in the low Prandtl number case.

%%%%%%%%%%%%%%%%%%%%%%%%%%%%%%%%%%%%%%%%%%%%%%%%%%%%%
%  5. The boundary layer ratio problem
%%%%%%%%%%%%%%%%%%%%%%%%%%%%%%%%%%%%%%%%%%%%%%%%%%%%%%%

\section{\label{section5}The boundary layer ratio problem}

In Boussinesq convection, there is a symmetry about the mid-plane which means that the top
and bottom boundary layers have the same thickness and structure, and the temperature of the bulk
interior is midway between that of the boundaries. In anelastic convection, this symmetry no longer holds,
and the top and bottom boundary layers can be very different, and the entropy of the bulk interior is
significantly different from $\Delta S / 2$. This raises the question of how the ratios of the thicknesses
of the top and bottom boundary layers, the ratio of the bulk horizontal velocities just outside the boundary layers,
and the ratio of the entropy jumps across the layers are actually determined. We assume the incompressible boundary layer case holds throughout this section.

We write the ratio of the entropy jumps across the boundary layers as
\begin{equation}
r_s = \frac{\Delta S_T}{\Delta S_B}, \ \ \textrm{so} \ \ \Delta S_B=\frac{\Delta S}{1+r_s} \  \ 
\textrm{and the entropy in the bulk is} \quad \frac{r_s}{1+r_s} \Delta S,
\label{eq:5_1}
\end{equation}
and the ratio of the anelastic temperature jumps across the boundary layers as
\begin{equation}
r_T = \frac{\Delta T_T}{\Delta T_B}.
\label{eq:5_2}
\end{equation}
We define the ratio of the velocities at the edge of the boundary layers as
\begin{equation}
r_u = \frac{U_T}{U_B}.
\label{eq:5_3}
\end{equation}
The last important ratio is the ratio of the thicknesses of the boundary layers. In general the viscous
and thermal boundary layers will be of different thickness, but here we start with the thermal boundary layers which have thicknesses at the top and bottom of $\delta^{th}_T$ and $\delta^{th}_B$ with
ratio
\begin{equation}
r_\delta = \frac{\delta^{th}_T}{\delta^{th}_B}.
\label{eq:5_4}
\end{equation}
We have four unknown ratios, so we need four equations to determine them. Our first equation 
uses the fact that the heat flux passing through the bottom boundary  must equal the heat flux 
passing through the top boundary. Since this heat flux is entirely by conduction close to the boundary,
 \begin{equation}
-k \frac{dT}{dz}\vert_i \sim k \frac{\Delta T_i}{\delta^{th}_i} \Rightarrow r_\delta = r_T.   
\label{eq:5_5}
\end{equation}
We can use the balance of advection and diffusion in the boundary layers exploited in \S4
to obtain another equation relating the boundary layer ratios. In \S4 we saw that the ratio of the
thermal boundary layer thicknesses was the same as the ratio of the viscous boundary layer thicknesses, 
 \begin{equation}
 \frac{\delta^{th}_T}{\delta^{th}_B} =  \frac{\delta^{\nu}_T}{\delta^{\nu}_B} = r_\delta,   
\label{eq:5_6}
\end{equation}
so we use the viscous boundary balance equation (\ref{eq:4_2}) to estimate 
\begin{equation}
\frac{\rho_B U_B^2 }{d} \sim \frac{\mu U_B}{{\delta^{\nu}_B}^2}, \quad
\frac{\rho_T U_T^2 }{d} \sim \frac{\mu U_T}{{\delta^{\nu}_T}^2} \quad \Rightarrow r_u {r_\delta}^2 \sim \frac{\rho_B}{\rho_T} = \Gamma^{m}.
\label{eq:5_7}
\end{equation}

We now need an equation for the ratio of bulk large scale flow velocities at the top and bottom of the layer, $r_U$.
We consider first the case where the viscous dissipation occurs primarily in the boundary layers, which is likely
to be true in numerical simulations with no-slip boundaries. Since the entropy production occurs in the boundary layers
and is relatively small in the interior, and entropy diffusion is small in the bulk interior, the convected 
entropy flux $\langle \bar \rho u_z s \rangle_h$
is approximately constant in the bulk interior. We now multiply the equation of motion (\ref{eq:2_1}) by ${\bar \rho} {\bf u}$ and average
over the bulk interior, ignoring the small viscous term in the bulk, to get
 \begin{equation}
\frac{1}{2}\frac{\partial}{\partial z}\left(\bar{\rho}\left\langle u_{z}u^{2}\right\rangle _{h}\right)
\approx  -\frac{\partial}{\partial z} \langle u_z p \rangle_h + \frac{g}{c_{p}}\left\langle {\bar \rho} u_{z}s\right\rangle _{h} 
\label{eq:5_8}.
\end{equation}
Near the boundary layers, the pressure term $p$ will be approximately constant as shown in (\ref{eq:2_23}),
and since $\langle u_z \rangle_h =0$, the  term involving $\langle u_z p \rangle_h$ will be small there, and we ignore it.
Since we expect $\langle \bar \rho u_z s \rangle_h$ to be approximately the same just outside the two boundary layers,  
\begin{equation}
\frac{\partial}{\partial z}\left(\bar{\rho}\left\langle u_{z}u^{2}\right\rangle _{h}\right) \vert_T \approx
\frac{\partial}{\partial z}\left(\bar{\rho}\left\langle u_{z}u^{2}\right\rangle _{h}\right) \vert_B,
\label{eq:5_9}
\end{equation}
where here $T$ and $B$ refer to conditions at the top of the bulk and the bottom of the bulk respectively.
In the turbulent bulk interior (unlike the boundary layers), we expect the three components of velocity to have
similar magnitudes. It remains to estimate the length-scale associated with the $z$-derivative,
and this is perhaps the most uncertain part of the analysis. Astrophysical mixing length theory 
uses the pressure scale height, or a multiple of the pressure scale height, as the mixing length for
the vertical length scale. Since we are only interested in the top and bottom ratios here, our results
are independent of what multiple of the scale height is chosen. Some support for the mixing length
idea can be derived from Kessar \textit{et al.} (2019), which shows that 
convective length scales decrease in the bulk towards the top of the layer. We also note that because we
are only concerned with ratios, it doesn't matter whether the pressure scale height or the density scale height 
is used. Adopting the pressure scale height,
 \begin{equation}
H_T = \frac{ d}{(m+1) ( \Gamma-1)}, \quad H_B = \frac{ \Gamma d}{ (m+1) (\Gamma-1)} \quad \Rightarrow \quad \frac{H_T}{H_B} = \Gamma^{-1}
\label{eq:5_10} 
\end{equation}
so that equation (\ref{eq:5_9}) gives
 \begin{equation}
\frac{\rho_T u_T^3}{H_T} \sim \frac{\rho_B u_B^3}{H_B} \Rightarrow r_u^3 \sim \Gamma^{m-1} \Rightarrow r_u \sim \Gamma^{\frac{m-1}{3}}.
\label{eq:5_11} 
\end{equation}

From the incompressible boundary layer equation (\ref{eq:2_25}) we have $r_s = \Gamma r_T$, so with the other three ratio equations   (\ref{eq:5_5}),
(\ref{eq:5_7}) and (\ref{eq:5_11}) we have
\begin{equation*}
\refstepcounter{equation}
r_s = \Gamma r_T, \quad r_\delta= r_T, \quad r_U {r_\delta}^2 = \Gamma^{m}, \quad {r_u} = \Gamma^{\frac{m-1}{3}},
\eqno{(\theequation \textit{a,b,c,d})}  \label{eq:5_12}
\end{equation*}
with solution
\begin{equation}
r_u = \Gamma^{\frac{m-1}{3}},  \quad r_\delta = \Gamma^{\frac{2m+1}{6}}, \quad r_s = \Gamma^{\frac{2m+7}{6}}, \quad r_T = \Gamma^{\frac{2m+1}{6}}.
\label{eq:5_13}
\end{equation}
In the case where $m=3/2$, appropriate for ideal monatomic gas,
\begin{equation}
r_u = \Gamma^{\frac{1}{6}},  \quad r_\delta = \Gamma^{\frac{2}{3}}, \quad r_s = \Gamma^{\frac{5}{3}}, \quad r_T = \Gamma^{\frac{2}{3}}.
\label{eq:5_14}
\end{equation}
Since $\Gamma$ is always greater than 1 and can be large, we see that the entropy in the bulk is much closer
to the entropy of the bottom boundary than to the entropy of the top boundary. The top boundary layer is thicker
than the bottom boundary layer. The challenge for numerical simulations is to get to sufficiently high
Rayleigh number that the top boundary layer is truly thin, as required by our asymptotic analysis. The ratio of
the bulk velocities at the top and bottom, is only weakly dependent on $\Gamma$, so again rather large values of $Ra$ 
are required to establish the asymptotic trend.   

Note that in deriving equation (\ref{eq:5_7}) we assumed that the horizontal length scale for advection
along the boundary layer was $d$, as did Grossmann \& Lohse (2000). We found that choosing the 
vertical length scales in the bulk to be based on the pressure scale height in  equation (\ref{eq:5_10}) agreed reasonably with the numerics, see \S\,7 below, so
a natural question is whether the horizontal length scale near the top boundary becomes less than $d$ when the layer is strongly stratified. The picture from our numerics is
somewhat mixed, and is discussed further in \S\,7 below. For moderate stratification, 
the numerics suggest the boundary layers do appear to extend to $d$ at both the top and bottom boundary; including
a factor $\Gamma$ in the horizontal length scales in the boundary layers gives poorer agreement with numerical
estimates of the boundary layer thickness ratio. However, at the largest values of $\Gamma$ we did find a departure from the (\ref{eq:5_14}) scalings
which could be accounted for by some reduction in the horizontal length scale near the top boundary. 

In the case where the viscous dissipation is mainly in the bulk, which happens at low $Pr$ (Grossmann \& Lohse, 2000)
the equations (\ref{eq:5_1}-\ref{eq:5_7}) still hold, but the argument  for equation (\ref{eq:5_11}) breaks down
because the entropy flux is no longer approximately constant in the bulk because viscous dissipation in the bulk is no longer
negligible. This case is discussed in Appendix B.

%%%%%%%%%%%%%%%%%%%%%%%%%%%%%%%%%%%
%  6. The Nusselt number and Reynolds number scaling laws
%%%%%%%%%%%%%%%%%%%%%%%%%%%%%%%%%%%%

\section{\label{section6}The Nusselt number and Reynolds number scaling laws}

When the boundary layers are thin, the overall entropy balance reduced to (\ref{eq:3_16}),
\begin{equation}
 \frac{F^{super} \Delta {\bar T}}{T_B T_T}  \sim
   \int_0^d \frac{\mu}{\bar T} \left\langle q \right\rangle_h \, dz.  
\label{eq:6_1}
\end{equation}
If the dissipation is mainly in laminar boundary layers, the $z$-derivatives of the horizontal velocities will dominate terms in the expression for $q$, so
\begin{equation}
q=(\partial_{j}u_{i})(\partial_{j}u_{i})+\frac{1}{3}\left(\nabla\cdot\mathbf{u}\right)^{2}
\approx  \left(\frac{\partial u_x}{\partial z}\right)^2 + \left(\frac{\partial u_y}{\partial z}\right)^2  \sim \frac{U_i^2}{(\delta_i^{\nu})^2},\label{eq:6_2}
\end{equation}
in these layers. So integrating over the boundary layers of thickness $\delta_i^{\nu}$ and assuming
$T$ varies little in these layers,
\begin{equation}
 \frac{F^{super} \Delta {\bar T}}{T_B T_T}  \sim
   \frac{\mu U_B^2}{T_B \delta_B^{\nu}} + \frac{\mu U_T^2}{T_T \delta_T^{\nu}}
= \frac{\mu U_B^2}{T_B \delta_B^{\nu}}\left(1+ \frac{\Gamma r_u^2}{r_\delta}\right).
\label{eq:6_3}
\end{equation}
where we have used the ratios (\ref{eq:5_3}), (\ref{eq:5_4}) and (\ref{eq:2_8}g). Now we
can write the superadiabatic flux in terms of the thermal boundary layer thicknesses, using
(\ref{eq:2_14}), (\ref{eq:2_26}), (\ref{eq:2_25}), (\ref{eq:5_1}) and (\ref{eq:2_18})
to get
\begin{equation}
 F^{super} =
   \frac{k \Delta T_B}{\delta_B^{th}} = \frac{k T_B \Delta S}{c_p (1+r_s) \delta_B^{th}}.
\label{eq:6_4}
\end{equation}
Inserting this into (\ref{eq:6_3}) and using the definition (\ref{eq:2_16}) for the Rayleigh number,
the entropy balance equation can be written
\begin{equation}
 \frac{k^2 Ra \Gamma}{c_p^2 (1+r_s) d^2 \rho_B^2} \frac{\delta_B^{\nu}}{\delta_B^{th}} \sim U_B^2 \left( 1+ \frac{\Gamma r_u^2}{r_\delta}\right).
\label{eq:6_5}
\end{equation}
Now we introduce the Reynolds number near the bottom boundary 
\begin{equation}
 Re_B = \frac{\rho_B U_B d}{\mu},
\label{eq:6_6}
\end{equation}
noting that the Reynolds number near the top, $Re_T$,  is given by $r_u Re_B$. We also use the definition of  the
Prandtl number,  (\ref{eq:4_1}), to write  (\ref{eq:6_5}) as
\begin{equation}
 \frac{Ra \Gamma}{Pr^2 (1+r_s) } \frac{\delta_B^{\nu}}{\delta_B^{th}} \sim Re_B^2 \left( 1+ \frac{\Gamma r_u^2}{r_\delta}\right).
\label{eq:6_7}
\end{equation}
The entropy balance equation has thus given us a relation between the Reynolds number and the Rayleigh number, which
is similar to that of regime I of Grossmann \& Lohse (2000) but with additional factors of $\Gamma$. In the high Prandtl number limit
  applying (\ref{eq:4_4}) gives
\begin{equation}
 Re_B \sim Ra^{1/2} Pr^{-5/6} \Gamma^{1/2} (1+r_s)^{-1/2} \left( 1+ \frac{\Gamma r_u^2}{r_\delta}\right)^{-1/2},
\label{eq:6_8}
\end{equation}
while in the low Prandtl number case we get using  (\ref{eq:4_5})
\begin{equation}
 Re_B \sim Ra^{1/2} Pr^{-3/4} \Gamma^{1/2} (1+r_s)^{-1/2} \left( 1+ \frac{\Gamma r_u^2}{r_\delta}\right)^{-1/2}.
\label{eq:6_9}
\end{equation}

We now use the viscous boundary layer balance between advection and diffusion,
 (\ref{eq:4_2}), $\rho_B U_B/d \sim \mu /  (\delta^{\nu}_B)^2$,  to obtain a balance 
between $Nu$ and $Re_B$. The boundary layer balance becomes
\begin{equation}
 Re_B \sim \left( \frac{d}{\delta_B^{\nu}} \right)^2, \ \textrm{but} \  Nu = \frac{d}{\delta_B^{th}} \frac{\Gamma \ln \Gamma}{(1+r_s)(\Gamma-1)}
\label{eq:6_10}
\end{equation}
using the expression (\ref{eq:2_27}) for the Nusselt number together with  (\ref{eq:2_12}) and (\ref{eq:5_1}).
Eliminating $d/\delta_B^{th}$ between these,
\begin{equation}
Re_B^{1/2}  = \frac{\delta_B^{th}}{\delta_B^{\nu}} \frac{Nu (\Gamma - 1)(1+r_s)}{\Gamma \ln \Gamma}.
\label{eq:6_11}
\end{equation}
As above, the ratio of the boundary layer thicknesses can be evaluated in terms of the Prandtl number, and
(\ref{eq:6_11}) allows us to eliminate $Re_B$ from (\ref{eq:6_7}) to obtain the Nusselt number as a function of Rayleigh number,
\begin{equation}
 \frac{Ra \Gamma}{Pr^2 (1+r_s) } \left( \frac{\delta_B^{\nu}}{\delta_B^{th}} \right)^5 \sim 
\left( \frac{ Nu (\Gamma-1)(1+r_s)}{\Gamma \ln \Gamma} \right)^4 \left( 1+ \frac{\Gamma r_u^2}{r_\delta}\right).
\label{eq:6_12}
\end{equation}
At large $Pr$, (\ref{eq:4_4}) gives 
the Nusselt number -- Rayleigh number scaling in the form
\begin{equation}
 Nu \sim Ra^{1/4} Pr^{-1/12} \frac{ \Gamma^{5/4} \ln \Gamma}{ \Gamma-1} 
(1+r_s)^{-5/4}  \left( 1+ \frac{\Gamma r_u^2}{r_\delta}\right)^{-1/4},
\label{eq:6_13}
\end{equation}
while at low $Pr$  (\ref{eq:4_6}) gives
\begin{equation}
 Nu \sim Ra^{1/4} Pr^{1/8} \frac{ \Gamma^{5/4} \ln \Gamma}{ \Gamma-1} 
(1+r_s)^{-5/4}  \left( 1+ \frac{\Gamma r_u^2}{r_\delta}\right)^{-1/4}.
\label{eq:6_14}
\end{equation}
If we accept the ratio scalings derived in \S5, in the case of a monatomic ideal gas, $\gamma=5/3, m=3/2$, we can write
\begin{equation}
 (1+r_s)^{-5/4}  \left( 1+ \frac{\Gamma r_u^2}{r_\delta}\right)^{-1/4} = 
\left(1+\Gamma^{5/3}\right)^{-5/4}  \left( 1+\Gamma^{2/3}\right)^{-1/4},
\label{eq:6_15}
\end{equation}
so as $\Gamma$ becomes large, $Nu$ decreases as $\ln \Gamma \Gamma^{-2}$. So we expect the Nusselt number, the dimensionless measure of the heat transport, to be considerably smaller when the layer is strongly compressible, i.e. when $\Gamma$ is large and there are many density scale heights in the layer at a given $Ra$ and $Pr$. Analogous results for the case where the dissipation is mainly in the bulk rather than in the boundary layers, as can happen at low $Pr$, are given in Appendix B.  

In the Boussinesq limit, $\Gamma$ is close to unity and $\theta =(\Gamma-1)/\Gamma$ is small, so $\ln \Gamma / (\Gamma-1) \to 1$ and $\rho_B \to \rho_T$ and $T_B \to T_T$. Equation (\ref{eq:2_1}) reduces to the usual Boussinesq equation with $s/c_p$ replaced by $\alpha T$, where for a perfect gas $\alpha=1/{\bar T}$ is the coefficient of expansion,
consistent with (\ref{eq:2_25}). The total jump of entropy across the layer, $\Delta S$, is replaced by the total temperature jump $\Delta T = \Delta S / \alpha c_p$ so the Rayleigh number (\ref{eq:2_16}) reduces to the familiar form
$Ra = g \alpha \Delta T d^3 / \kappa \nu$ where $\kappa=k/{\bar \rho} c_p$ and $\nu = \mu / {\bar \rho}$ are the thermal diffusivity and kinematic viscosity respectively. These are both constant in the Boussinesq limit, and the ratios
$r_u$, $r_{\delta}$ and $r_s$ all go to unity, see \S 5.  Our scaling laws  (\ref{eq:6_8}), (\ref{eq:6_9}), 
 (\ref{eq:6_13}) and (\ref{eq:6_14}) all reduce to those of regimes $I_u$ and $I_l$ of Grossmann \& Lohse (2000). Grossmann \& Lohse also give 
suggested prefactors for their scaling laws in table 2 of their paper, and since our anelastic scaling laws reduce to theirs in the Boussinesq limit, our prefactors should in theory be consistent with theirs. In practice, the prefactors depend on the aspect ratio of the experiments (or numerical experiments)
used to determine them (see e.g. Chong \textit{et al.}, 2018). For the case of the high Prandtl number regime $I_u$,
their values were $Nu \approx 0.33 Ra^{1/4} Pr^{-1/12}$ and $Re \approx 0.039 Ra^{1/2} Pr^{-5/6}$, so
 (\ref{eq:6_13}) becomes
\begin{equation}
 Nu \approx C_{Nu} Ra^{1/4} Pr^{-1/12} \frac{ \Gamma^{5/4} \ln \Gamma}{ \Gamma-1} 
(1+r_s)^{-5/4}  \left( 1+ \frac{\Gamma r_u^2}{r_\delta}\right)^{-1/4},  \quad C_{Nu}=0.93.
\label{eq:6_16}
\end{equation}
In the low $Pr$ limit where (\ref{eq:6_14}) applies,  regime $I_l$ of Grossmann \& Lohse (2000), they suggest a prefactor corresponding to $C_{Nu} = 0.76$ rather than 0.93. For the Reynolds number, (\ref{eq:6_8})  becomes
\begin{equation}
 Re_B \approx  C_{Re} Ra^{1/2} Pr^{-5/6} \Gamma^{1/2} (1+r_s)^{-1/2} \left( 1+ \frac{\Gamma r_u^2}{r_\delta}\right)^{-1/2}, \quad C_{Re}=0.078.
\label{eq:6_17}
\end{equation} 
There is some uncertainty about the prefactor $C_{Re}$, discussed in \S7 below.  
Prefactors in the case $I_l$ and in the case where dissipation is mainly in the bulk, their case $II_l$, (see Appendix B)
can also be found.

%%%%%%%%%%%%%%%%%%%%%%%%%%%%%%%%%%%%%%%%%%%%%%%%%%%%%
%  7. The numerical results and discussion
%%%%%%%%%%%%%%%%%%%%%%%%%%%%%%%%%%%%%%%%%%%%%%%%%%%%%%%

\section{\label{section7}The numerical results and discussion}

%
%Figure 3
%
\begin{figure}
\begin{centering}
\vspace{3mm}
\includegraphics[scale=0.33]{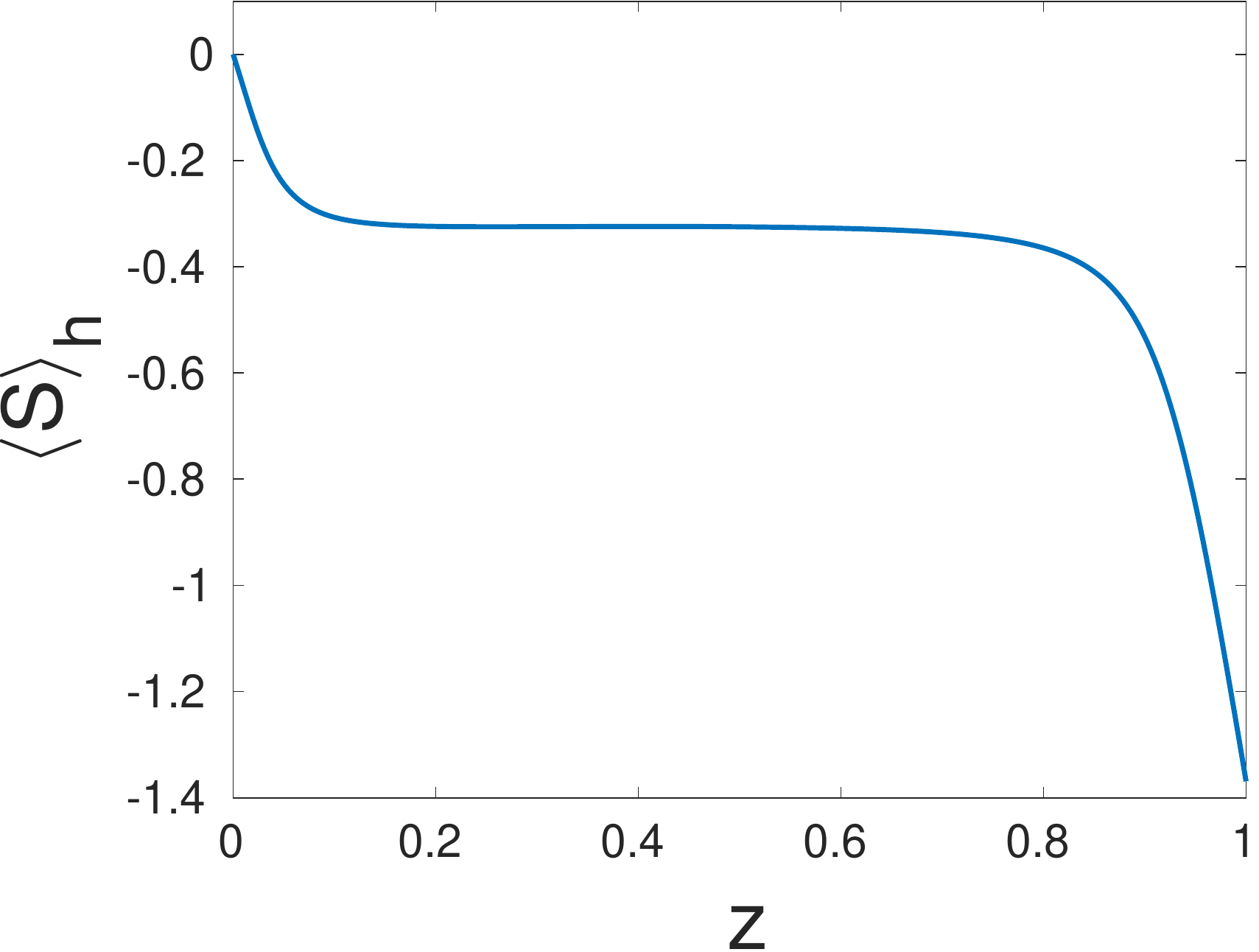} \hspace{5mm}  \includegraphics[scale=0.33]{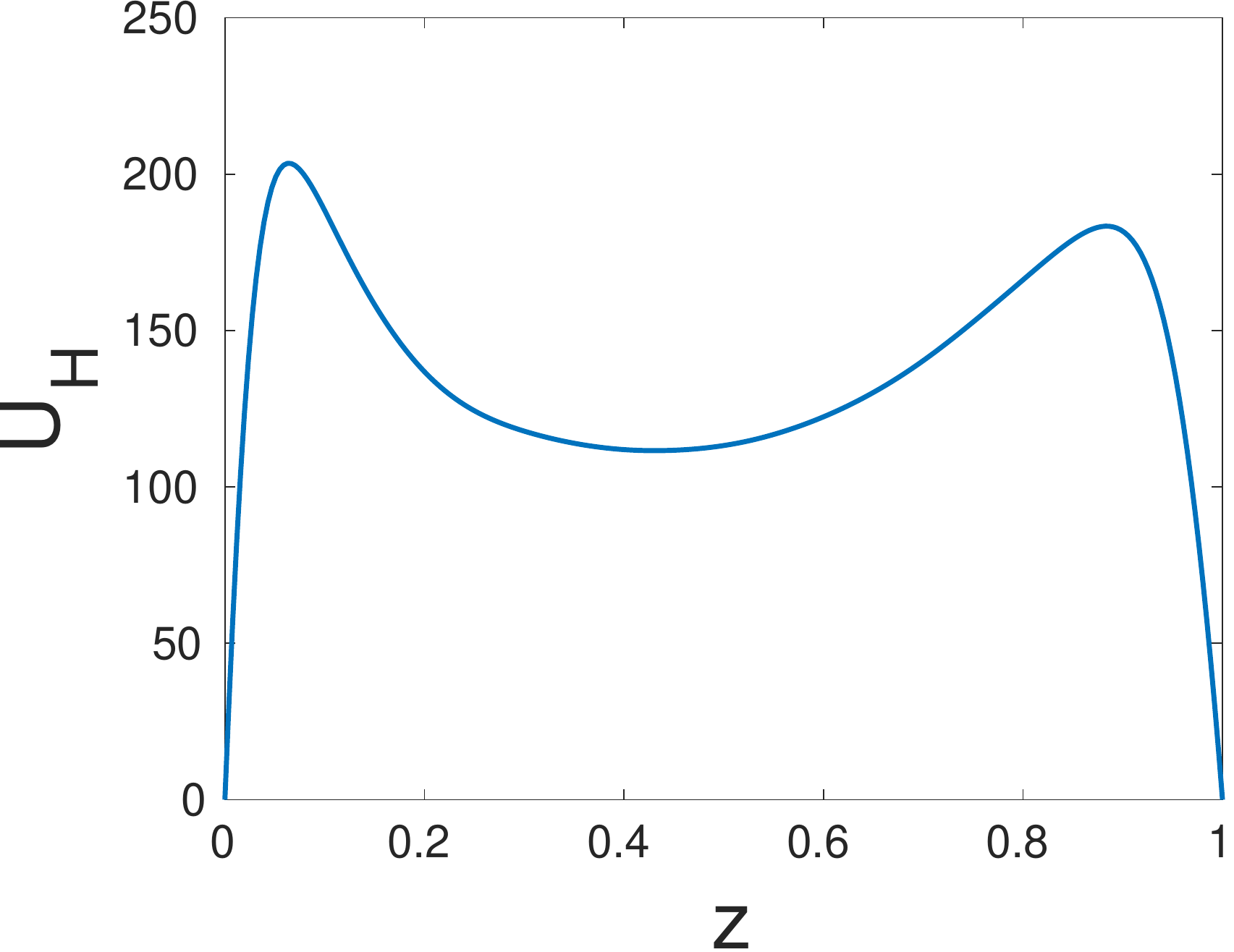}

\vspace{2mm}

\includegraphics[scale=0.33]{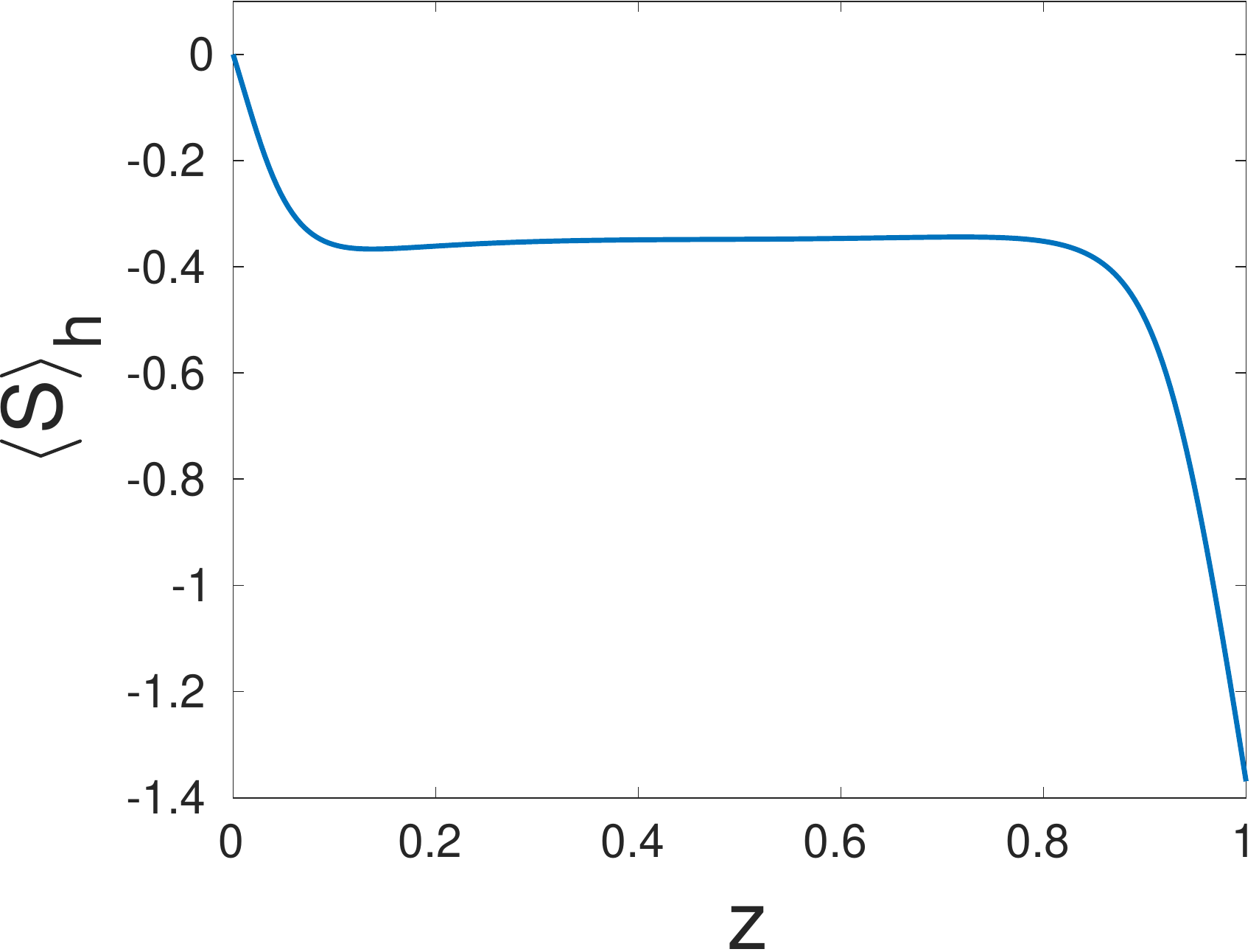} \hspace{5mm}  \includegraphics[scale=0.33]{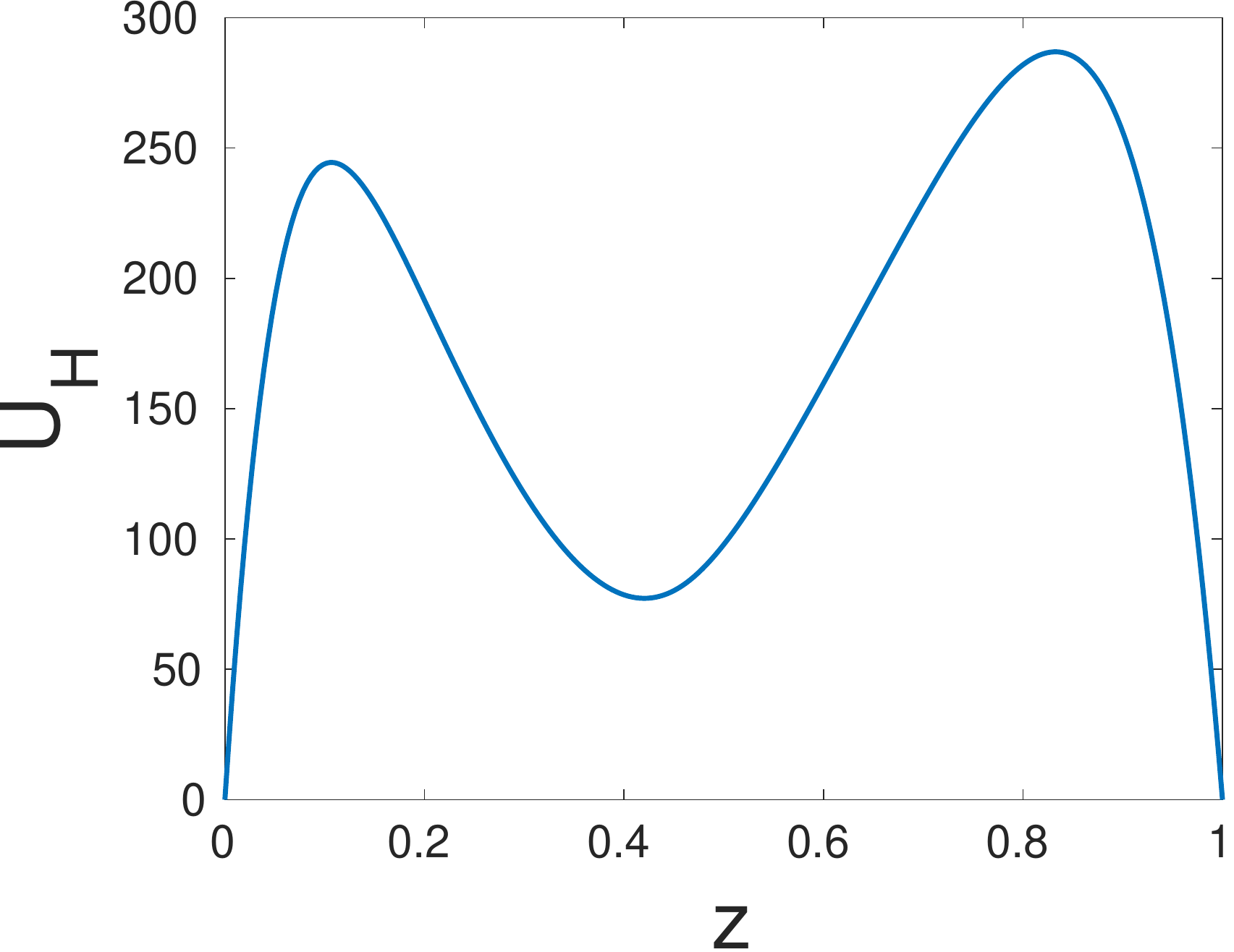}

\vspace{2mm}

\includegraphics[scale=0.33]{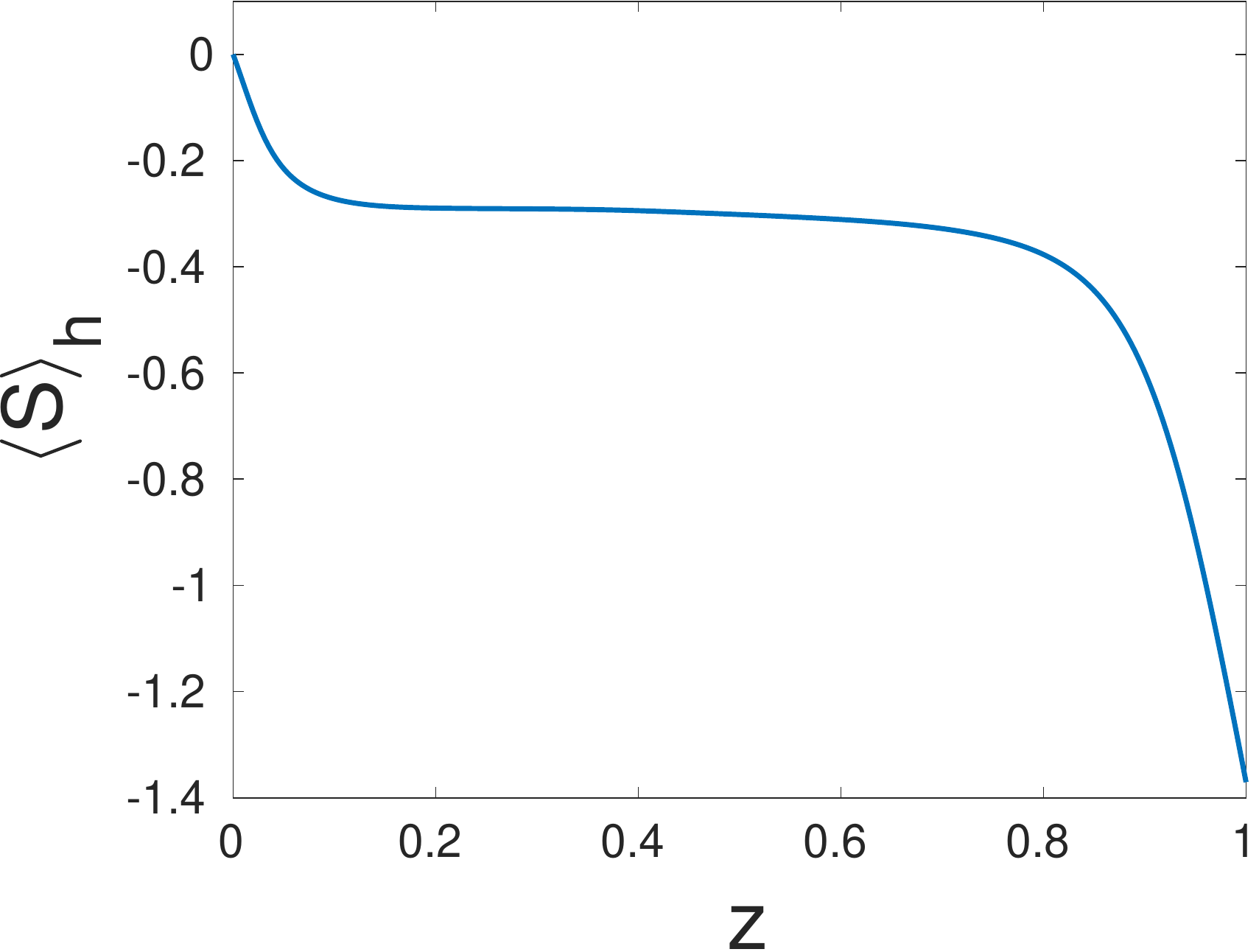} \hspace{5mm}  \includegraphics[scale=0.33]{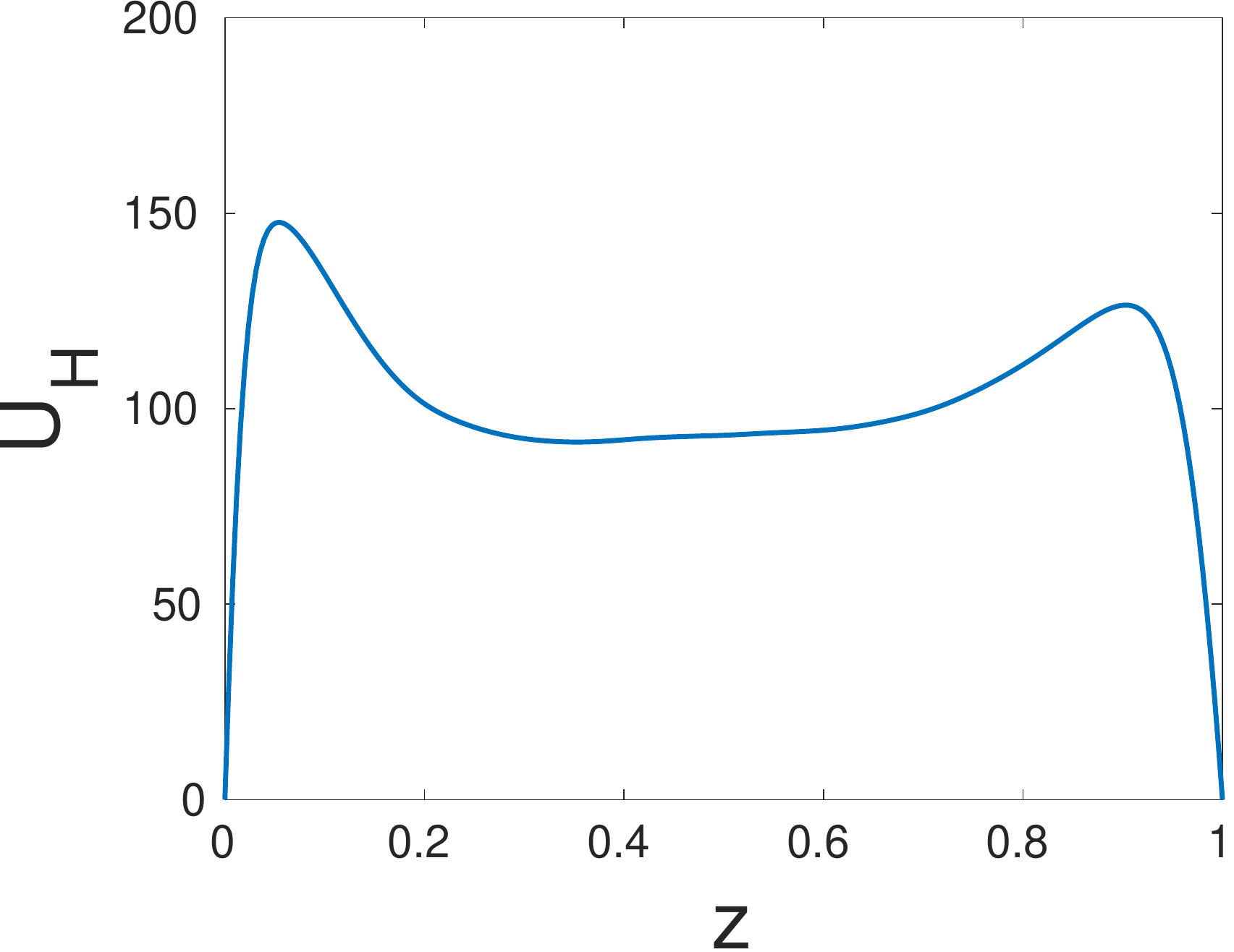}
\par\end{centering}
\vspace{-137mm}
\hspace{5mm} (a) \hspace{60mm} (b)

\vspace{43mm}
\hspace{5mm} (c) \hspace{60mm} (d)

\vspace{43mm}
\hspace{5mm} (e) \hspace{60mm} (f)

%\vspace{13mm}
%\hspace{-3mm} $<s>_h$ \hspace{57mm} $U_h$ 

%\vspace{20mm}
%\hspace{31mm} z \hspace{62mm} z

\vspace{40mm}

\caption  {Horizontally averaged entropy (in units of $c_p$) and horizontal mean velocity profiles (Peclet number units) from the numerical simulations for $\Gamma=1.9438$, $Ra=10^6$.  (a) Entropy profile at $Pr=1$.
 (b)  Horizontal velocity  profile at $Pr=1$. (c)  Entropy profile at $Pr=10$.
 (d)  Horizontal  velocity  profile at $Pr=10$. (e)  Entropy profile at $Pr=0.25$. 
(f)  Horizontal velocity  profile at $Pr=0.25$.}
\label{fig3}
\end{figure}

%
%Figure 4 
%

\begin{figure}
\begin{centering}
\vspace{3mm}
%Data from /home/caj30/cajones/MOULOUD/X5_NS_PR1_0_RA3_106
% and /home/caj30/cajones/MOULOUD/X10_NS_PR1_0_RA6_106_512
\includegraphics[scale=0.33]{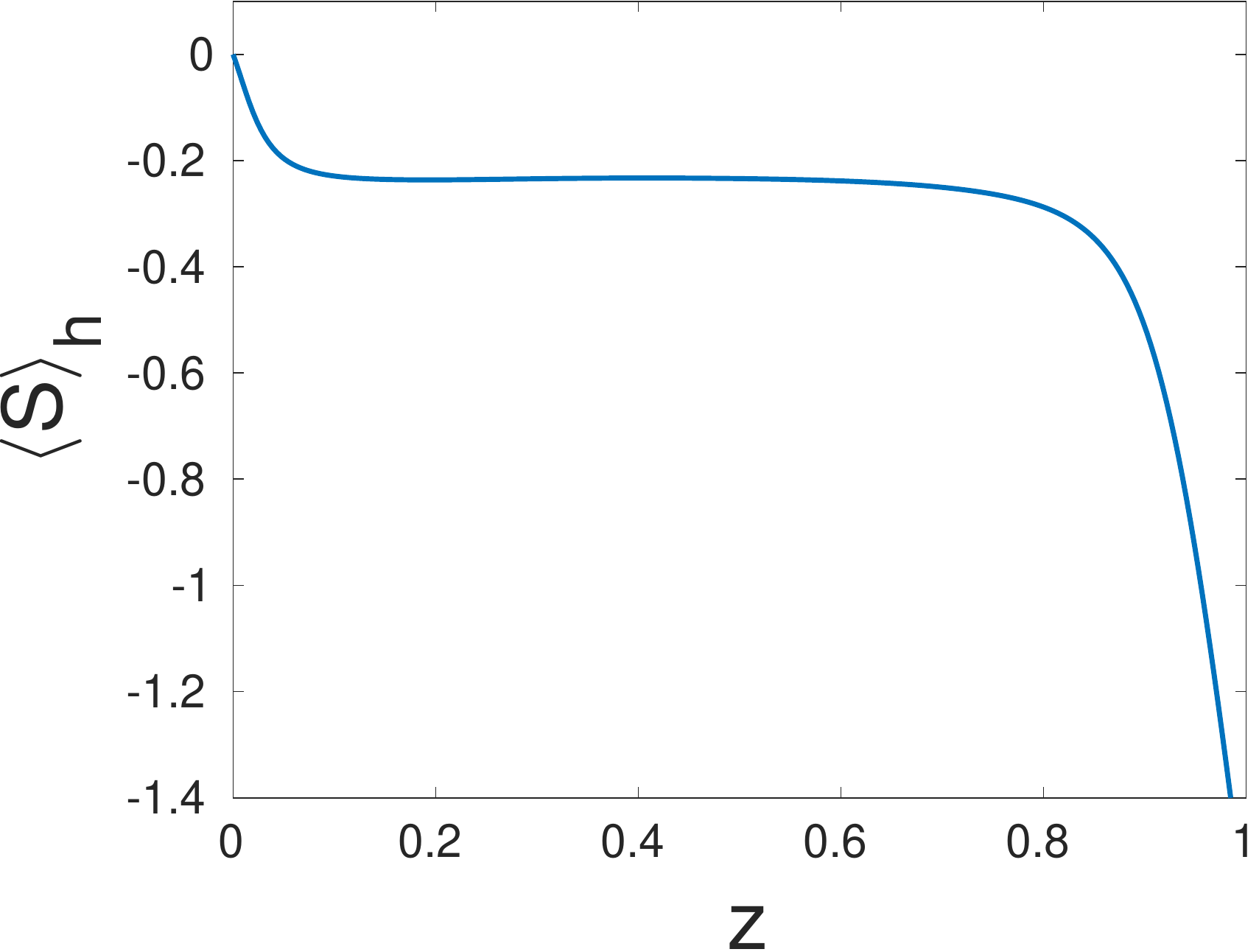} \hspace{5mm}  \includegraphics[scale=0.33]{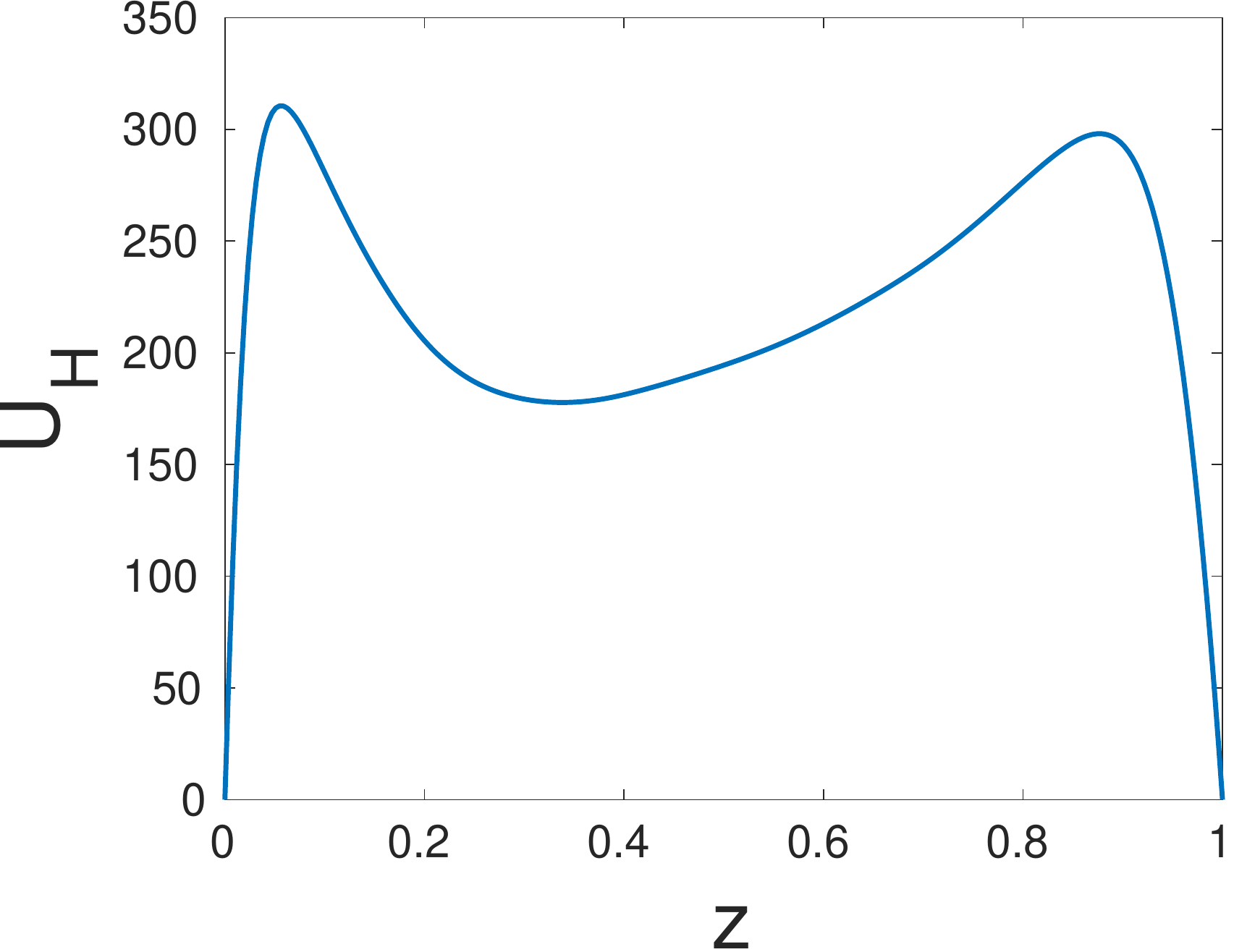}

\vspace{2mm}

\includegraphics[scale=0.33]{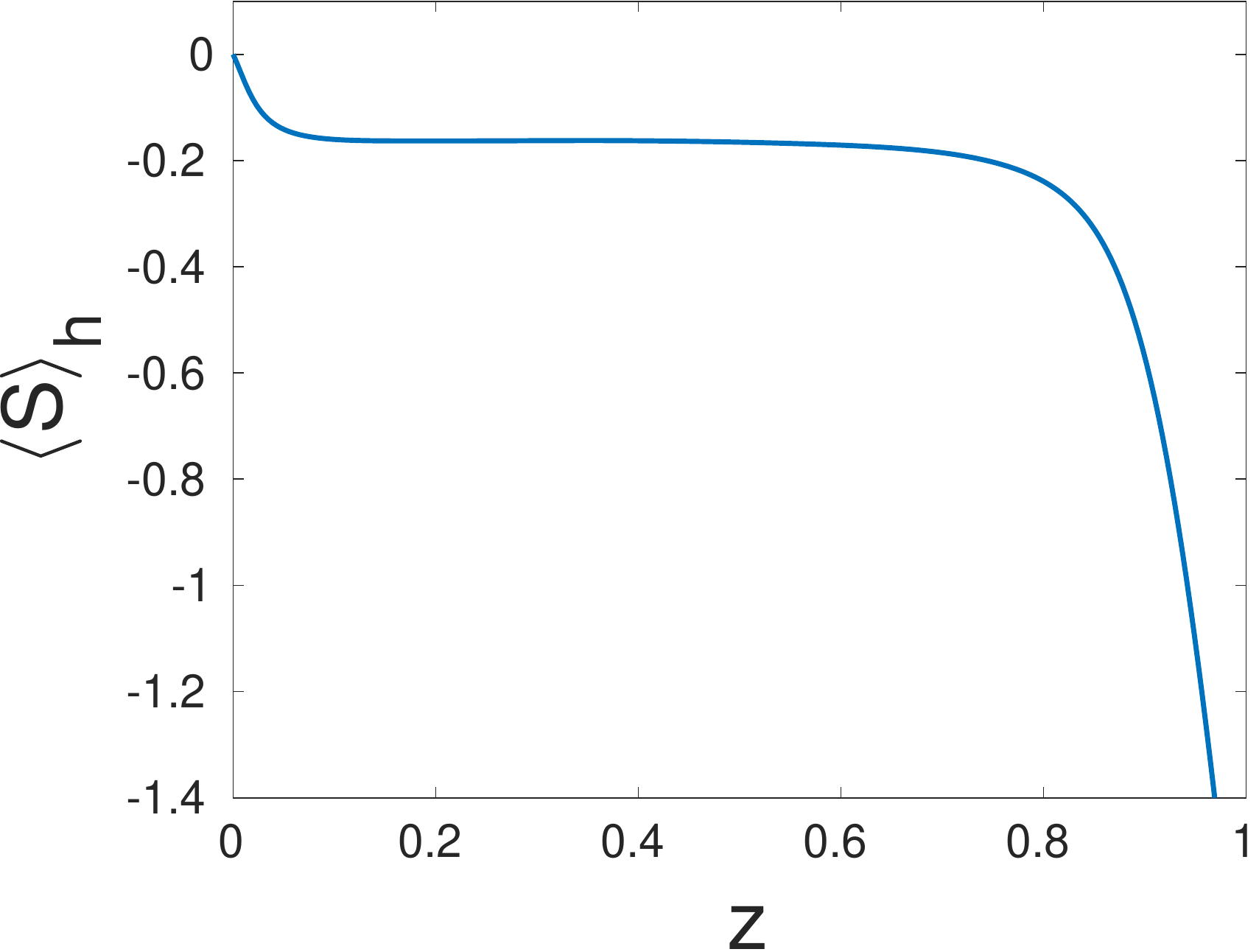} \hspace{5mm}  \includegraphics[scale=0.33]{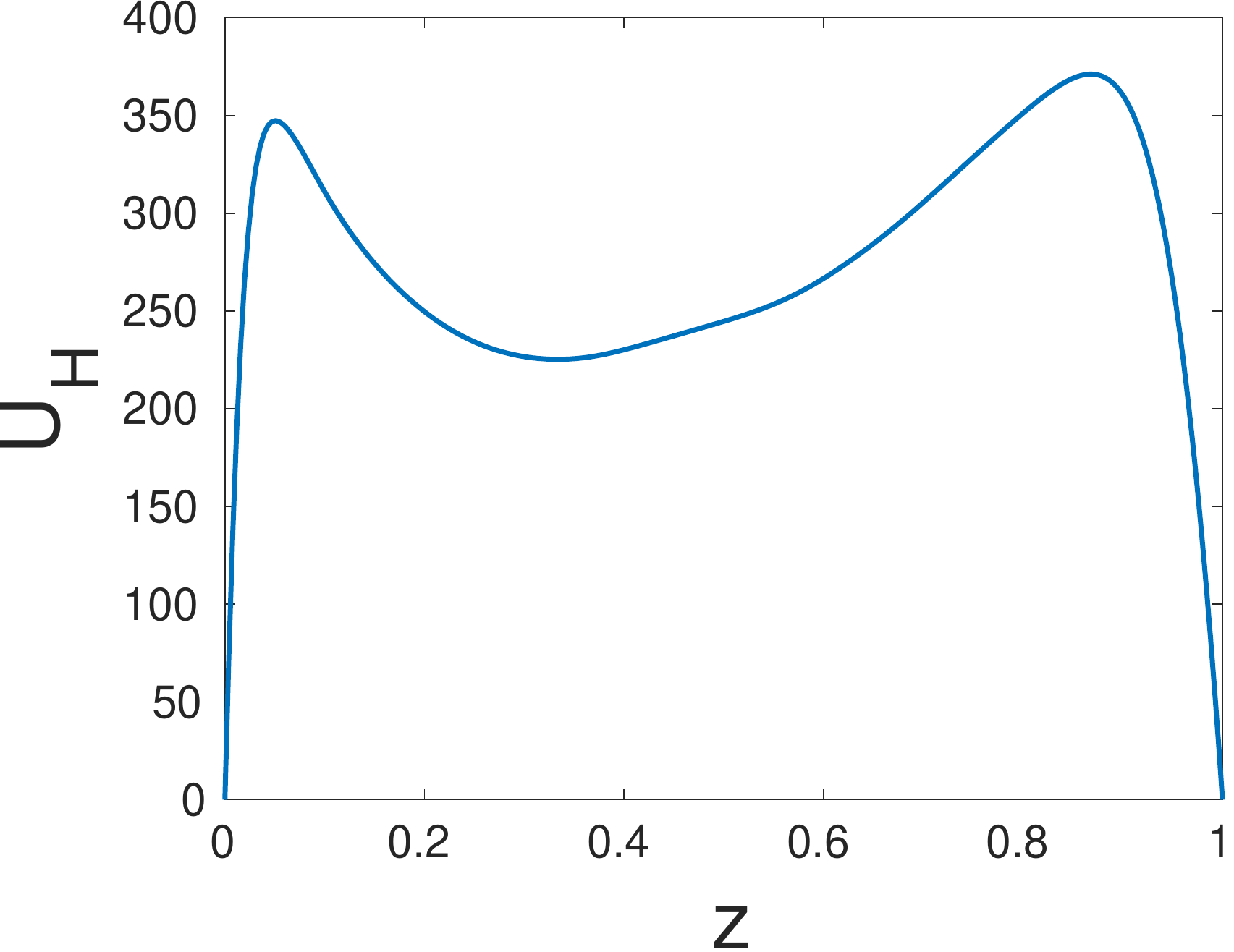}
\par\end{centering}

\vspace{-90mm}
\hspace{5mm} (a) \hspace{60mm} (b)

\vspace{43mm}
\hspace{5mm} (c) \hspace{60mm} (d)

%\vspace{20mm}
%\hspace{31mm} z \hspace{62mm} z

\vspace{40mm}

\vspace{2mm}

\caption  {Horizontally averaged entropy ${\langle}S{\rangle}_h$  and horizontal mean velocity $U_H$  profiles for (a,b) $\Gamma=2.924$, $Ra=3 \times 10^6$, $Pr=1$:
  (c,d) $\Gamma=4.6416$, $Ra=6 \times 10^6$, $Pr=1$.} 
\label{fig4}
\end{figure}

We have tested the theoretical predictions of our asymptotic theory using numerical simulations of high Rayleigh number plane layer
anelastic convection. The numerical code differs from the theory in one respect, as it uses entropy diffusion $k_s$ rather than temperature
diffusion in the heat equation, so 
\begin{equation}
\bar{\rho}\bar{T}\left[\frac{\partial s}{\partial t}+\mathbf{u}\cdot\nabla s \right]=k_s \nabla \cdot {\bar T} \nabla s +\mu\left[q+\partial_{j}u_{i}\partial_{i}u_{j}-\left(\nabla\cdot\mathbf{u}\right)^{2}\right], \label{eq:7_1}
\end{equation}
%\end{document}
where $k_s$ is constant, replaces (\ref{eq:2_3}). This simplifies the code because entropy is the only anelastic thermodynamic variable computed, and it can be justified in circumstances where
turbulent diffusion dominates laminar diffusion (Braginsky \& Roberts, 1995). Constant entropy boundary conditions were used in the code.
The energy balance equation (\ref{eq:3_5}) is only changed by replacing $-k d{ \langle T \rangle}_h /dz$  by    $-k_s {\bar T}  d{\langle s \rangle}_h /dz$.
In the entropy balance equation, entropy diffusion changes $S_{diff}$ to
\begin{eqnarray}
  S_{diff} =- {k_s}  \frac{\mathrm{d}}{\mathrm{d}z} \left\langle s \right\rangle_h \Big|_{z=0}
+ {k_s} \frac{\mathrm{d}}{\mathrm{d}z} \left\langle s \right\rangle_h \Big|_z  
- \int_0^z \frac{k_s \Delta {\bar T}}{{\bar T} d} \frac{\mathrm{d}}{\mathrm{d}z} \left\langle s \right\rangle_h \, dz .
\label{eq:7_2}
\end{eqnarray}   
Just as in the temperature diffusion case, the last two terms are negligible when the entropy boundary layers are thin, except
when $z$ is in the boundary layers, when the second term is significant. In the case when the viscous dissipation is primarily in the boundary
layers, the argument leading to (\ref{eq:5_9}) still holds, so the ratios satisfy (\ref{eq:5_13}) in the entropy diffusion case
as well as in the temperature diffusion case. It is therefore reasonable to compare with the numerical results for the entropy diffusion case
in the expectation that they will be reasonably close to the temperature diffusion case.

\begin{table}
 \begin{center}
\def~{\hphantom{0}}
%\small
\begin{tabular}{ccccccc}
%\centering\small
%\begin{tabular}{rrrrrrrll}
%\hline
 Run  & A1 & A2 & A3 & A4 & A5   \\[3pt]
        & B1 & B2 & C1 &  D1   &     D2 & D3   \\
\hline
 Ra & $10^6$  & $3 \times 10^6$ & $ 10^7$ & $3 \times 10^6$ & $ 6 \times 10^6$ \\
     & $ 10^6$ &  $3 \times 10^6$ & $ 10^6$  &    $ 10^6$  & $ 10^6$  & $10^6$ \\
 
 Pr   & $1$ &  $1$ &  $1$ &  $1$ & $1$ \\
       & $10$ & $10$ & $0.25 $   & $1$ & $10$   & $0.25$ \\
 
 $\Gamma$    & $1.9438$ & $1.9438$ & $1.9438$ & $2.924$ & $4.6416$ \\
                     & $1.9438$ & $4.6416$ & $1.9438$ &       $1$      &     $1$ & $1$           \\

 $\rho_b / \rho_T$    & $2.71$ & $2.71$ & $2.71$ & $5$ & $10$  \\
                               & $2.71$ & $10$ & $2.71$    &   $1$  & $1$ &  $1$ \\
 
    & &  &  & &   \\

 $r_\delta$ & $1.65 \pm 0.01$ & $1.70 \pm 0.01$ & $1.69 \pm 0.01$ & $2.05 \pm 0.01$ & $2.27 \pm 0.02$ \\
                & $1.48 \pm 0.01$ & $2.17 \pm 0.01$ & $2.05 \pm 0.04 $  &  1   &  1  & 1    \\
 
 $r_s$  & $3.21 \pm 0.02$ & $3.29 \pm 0.01$ & $3.31 \pm 0.01$  & $5.99 \pm 0.02$ & $10.62 \pm 0.03$ \\ 
           & $2.91 \pm 0.02$ & $10.0 \pm 0.02$ & $4.04 \pm 0.04$  &       1       &  1    & 1     \\
 
 $r_u$    & $0.92 \pm 0.01$ & $0.92 \pm 0.01$ & $0.93 \pm 0.01$ & $0.96 \pm 0.01$ & $1.06 \pm 0.02$ \\ 
             & $1.17 \pm 0.01$  & $1.38 \pm 0.01$ & $0.82 \pm 0.02$  &  1  &  1  & 1  \\
 
 $Nu$    & $5.90 \pm 0.02$ & $7.75 \pm 0.03$ &  $10.69 \pm 0.03$ & $5.35 \pm 0.02$ & $4.26 \pm 0.02$ \\
            & $6.18 \pm 0.02$ & $ 4.09 \pm 0.02$ & $ 5.00 \pm 0.04$ & $8.78 \pm 0.03$ &  $8.76 \pm 0.04$  & $8.17 \pm 0.04$ \\

 $U_T$    & $186 \pm 3$ &  $321 \pm 3$ & $570 \pm 3$ & $298 \pm 3$ & $ 368 \pm 3$ \\ 
              & $ 287 \pm 2$ & $410 \pm 3$ &   $ 125 \pm 4 $     &  $ 200 \pm 3 $  & $ 260 \pm 5$ & $149 \pm 3$   \\
 
 $U_B$   & $202 \pm 3$ & $348 \pm 3$ & $613 \pm 3$ & $311 \pm 3$ & $ 348 \pm 3 $ \\
            & $244 \pm 2$ & $ 298 \pm 3$ & $ 154 \pm 4 $ &  $ 200 \pm 3 $ & $260 \pm 5$ & $149 \pm 3$ \\

   & &  &  & &   \\
 
 $\Gamma^{2/3}$   & 1.557 & 1.557 & 1.557 & 2.045 & 2.783 \\ 
                              & 1.557 & 2.783 & 1.557  & 1  &  1 & 1 \\
 
 $\Gamma^{5/3}$   & 3.028 & 3.028 & 3.028 & 5.979 & 12.915 \\
                              & 3.028 & 12.915 & 3.028 & 1   & 1 & 1 \\ 
 
$\Gamma^{1/6}$    & 1.117 & 1.117 & 1.117 & 1.196 & 1.292 \\
                              & 1.117 & 1.292 & 1.117  &  1   &  1 & 1 \\ 

$Nu$-theory & 5.56 & 7.32 & 9.89 &  4.65 & 2.98  \\
                   &  5.55  & 2.50 & 5.18 & 8.78  & 8.76  & 8.17\\ 

$Nu$-nblr & 5.60 & 7.22 & 9.67 &  4.97 & 3.86  \\
                   &  5.63  & 3.11 & 4.37 & 8.78  & 8.76  & 8.17\\ 

$Pe_T$-theory & 194 & 336 & 614 & 307 & 376 \\  
                      & 252 & 345 & 144  & 200 & 260 & 149   \\

$Pe_B$-theory & 174 & 301 & 549 & 257 & 291 \\
                     & 226 & 267 & 129  &  200 & 260 & 149  \\

$Pe_T$-nblr & 177 & 306 & 559 & 283 & 361 \\  
                      & 256 & 358 & 118  & 200 & 260 & 149   \\

$Pe_B$-nblr & 192 & 332 & 601 & 295 & 341 \\
                     & 219 & 260 & 144  &  200 & 260 & 149  \\
 
\end{tabular}

 \caption{Data from the numerical runs all corresponding to $m=3/2$ polytropes. The first four rows are the input parameters.  $r_\delta$, $r_s$ and $r_u$ are the measured boundary layer ratios for each run. The velocities $U_T$ and $U_B$  are the local maxima at the edge of the boundary layers, measured in velocity units of $k/d \rho_B c_p$, i.e they are Peclet numbers based on the diffusivity at the base of the layer. The theoretical predictions for the  boundary layer ratios are given in the next three rows, see equation (\ref{eq:5_14}). The $Nu$-theory entries are based on (\ref{eq:6_16}) with the prefactors $C_{Nu}$ as given in the text, and the boundary ratios come from (\ref{eq:5_14}). The $Nu$-nblr entries also use
(\ref{eq:6_16}) with the same prefactors, but instead of using (\ref{eq:5_14}), the numerical boundary layer ratios (nblr) above are used. 
The $Pe_T$-theory and $Pe_B$-theory entries come from  (\ref{eq:6_8}) and (\ref{eq:6_9}). The prefactors used are not those of Grossmann \& Lohse (2000),
see (\ref{eq:6_17}), but those given in the text. Again, (\ref{eq:5_14}) is used to determine the boundary layer ratios. The $Pe_T$-nblr and $Pe_B$-nblr entries use the numerical boundary layer ratios. }
\label{table:cases}
%\tablenotetext{a}{Footnote text here.}
  \end{center}
\end{table}

The numerical code is described in Kessar \textit{et al.} (2019), though for that paper stress-free boundary conditions were applied, whereas no-slip boundaries where imposed in the runs described here. The unit of length is taken as $d$, the unit of time is  $\rho_B d^2 c_p / k$, the thermal diffusion time at the bottom of the layer.  The velocities are in units of $k/ \rho_B d c_p$, so they correspond to local Peclet numbers, where $Pe=Re Pr$.

All the runs have polytropic index $m=3/2$. The code assumes periodic boundary conditions in the horizontal $x$ and $y$ directions, with aspect ratio 2, that is the period in the horizontal directions is $2d$. Table 1 gives the parameters used in the eleven runs, which span a range of Prandtl numbers and are at Rayleigh numbers which are as large as is numerically feasible, bearing in mind the need to resolve the small scale structures that develop. The last three runs are
for Boussinesq cases, $\Gamma=1$, for comparison with the anelastic cases. The density stratification measured by $\Gamma$ varies over a moderate range only, because for the modest values of $Ra$ that are numerically accessible, large $\Gamma$ leads to a top boundary layer which is no longer thin, so our theory will no longer be valid. In figure 3, the entropy profiles (in units of $c_p$) and the horizontal velocity profiles are shown for the three runs A1, B1 and C1, and the profiles for  A4 and A5 are shown in figure 4. The entropy profiles are constructed by horizontal averaging and time averaging the vertical profiles. From (\ref{eq:2_12})  the entropy difference between the boundaries is $\Gamma \ln \Gamma / (\Gamma-1)$ and the constant is chosen so that it is zero at the bottom boundary $z=0$.

From figure 3 it is immediately apparent that the entropy is indeed rather constant in the well-mixed bulk interior, consistent with a key assumption of the theory. It is also clear that the jump in entropy across the top boundary layer is greater than that across the bottom boundary layer, and that the top entropy boundary layer is thicker than the bottom entropy boundary layer. This is consistent with the boundary layer ratios found in \S5.  The velocity profiles have local maxima near the boundaries, which gives a convenient definition for the top and bottom Reynolds numbers, $Re_B$ and $Re_T$, after converting Peclet numbers to Reynolds numbers using $Re Pr = Pe$. We note that there is no great difference between the top and bottom horizontal velocities, consistent with the weak scaling with $\Gamma$ found in (\ref{eq:5_11}). This result is slightly surprising, because astrophysical mixing length theory predicts faster velocities where the fluid is less dense, but in our problem the boundaries play an important role.  The low Prandtl number case, figures 3(e) and 3(f), has a slightly different entropy boundary layer profile from those of the $Pr=1$ and $Pr=10$ cases, with a more gradual matching on to the uniform entropy interior, which is particularly noticeable in the upper boundary layer. This suggests it may be necessary to go to higher $Ra$ at low $Pr$ before accurate agreement with a theory that assumes thin boundary layers can be obtained.  The cases shown in figure 4, together with figures 3(a) and 3(b), form a sequence at constant $Pr=1$ with increasing $\Gamma$. The most noticeable feature is that at larger $\Gamma$ the entropy of the mixed interior becomes close to that of the bottom boundary, so the boundary layer ratio $r_S$ increases rapidly, consistent with the prediction of  (\ref{eq:5_14}). Also notable is that the velocity ratio  of the maximum horizontal velocities, $r_U$, is never far from unity, again consistent with the weak scaling with $\Gamma$ in (\ref{eq:5_14}). As expected, the boundary layers become thicker at larger $\Gamma$, so that at fixed $Ra$ the thin boundary layer assumption  breaks down at large $\Gamma$.     

In table 1 we compare various results of the simulations against the theoretical predictions of \S5 and \S6. To evaluate the entropy boundary layer thicknesses for our numerical data we use  the definition  (\ref{eq:2_22}), so the gradients $d \langle S \rangle_h /dz$ at $z=0$ and $z=1$ were obtained by differentiating a cubic spline representation of the entropy, and the entropy jumps were obtained by averaging the entropy over the well mixed region, assuming constant entropy there. 
 The ratios of the top and bottom entropy thicknesses and entropy jumps are denoted by $r_{\delta}$  and $r_s$ in table 1. The velocity ratios at the top and bottom are denoted by $r_u$. We can compare these with the predicted ratios in (\ref{eq:5_14}).  We see that there is some variation in the numerical results, but they are roughly in agreement with the predicted results. Considering that the top boundary layer is not that thin, as can be seen in figures 3 and 4, these results are as good as can be expected. We also tested how well the equations leading to our boundary layer ratios compare individually with the numerical results. Equation  (\ref{eq:5_5}) expresses the fact that the heat flux is the same at the top and bottom, and together with the incompressible boundary layer equation (\ref{eq:2_25}),
gives $r_s= \Gamma r_{\delta}$, which agrees with our numerics rather well, to  the 1\% level. The viscous boundary layer balance, (\ref{eq:5_7}), has less accurate agreement with the numerics. All the runs at $\Gamma=1.9438$ had errors less than 10\% (except the low $Pr$ run where as we saw in figure 3 the boundary layer structure is slightly different), but the runs with density ratio 10, A5 and B2, have $r_{\delta}$
and $r_s$  large, but not quite as large as predicted by  (\ref{eq:5_14}). The likely explanation is that the horizontal length scale at the top boundary layer is getting smaller than at the bottom boundary, though not by as much as the factor $\Gamma$ predicted for the vertical length scale ratio. There is therefore some doubt as to whether it is correct to have $d$ as the horizontal length scale in the boundary layers when $\Gamma$ is large. Further research is needed to elucidate this issue. The velocity ratio equation (\ref{eq:5_11}) correctly predicts that the velocity ratio is always close to unity, but is less reliable at predicting whether it is above or below unity. However, the higher density ratio runs do have $r_u>1$, as predicted by (\ref{eq:5_11}).

We also evaluated the Nusselt number from the data, using
 \begin{eqnarray}
  Nu  =  - \frac{d}{c_p} \frac{d \langle S \rangle_h}{dz} \Big{\vert}_{z=0}  =  
 - \frac{d}{\Gamma c_p} \frac{d \langle S \rangle_h}{dz} \Big{\vert}_{z=d}, \label{eq:7_3}
\end{eqnarray} 
again determining the gradients from our spline representation.   
When the run has been integrated long enough, initial transients in the numerical run are eliminated and these two estimates of the Nusselt number become close, and the average value is used in table 1. Because the flow is turbulent, the Nusselt number fluctuates
continuously at about the 10\% level, so a long time average is used. The finite length of the run means there is a small uncertainty due
to the fluctuations not exactly cancelling, which we estimate as error bars in table 1.  

In table 1 we also give the value of the Nusselt number calculated by our theory. We used the Boussinesq runs, D1, D2 and D3 to determine the prefactors $C_{Nu}$ appropriate for each Prandtl number used. This gives $C_{Nu}=0.949$ for $Pr=10$,  $C_{Nu}=0.785$ for $Pr=1$ and $C_{Nu}=0.869$ for $Pr=0.25$. Note that in table 1 this means that for the Boussinesq runs D1, D2, and D3, the $Nu$-theory entries are the same as the actual $Nu$ entries by construction. None of these prefactors is very far from the values suggested by Grossmann \& Lohse (2000), who had $C_{Nu} = 0.93$ at large $Pr$ and $C_{Nu}=0.76$ at small Prandtl number, suggesting that the differences in aspect ratio and geometry only make relatively small changes to the Nusselt number. For consistency, we use these same prefactors in all runs. Since our main interest is in the compressible cases, we have not explored why the prefactor for the $Pr=1$ case is slightly lower than the other values of $C_{Nu}$.

We first consider the Nusselt number for the cases where the density ratio is only 2.71, runs A1, A2, A3, B1 and C1. We note that all the predicted values are not too far off the numerical values, though the predicted values are generally  a little lower than the actual numerical values. 
 Using the numerically calculated boundary layer ratios in formula (\ref{eq:6_16}) rather than the theoretically predicted ones gives the result $Nu$-nblr
in table 1. These are only available after the simulation is run, but they are helpful for testing whether small errors are due to slightly inaccurate boundary layer ratios, or whether the formula  (\ref{eq:6_16}) is inaccurate. For the density ratio 2.71 runs with $Pr \ge 1$, the boundary layer ratios were close to the predicted values, so not surprisingly, $Nu$-nblr are not significantly better than $Nu$-theory. We conclude that the under-prediction of the Nusselt number in these cases, which is less than about the 10\% level, is due to the viscous dissipation not being completely in the boundary layers, as required by the theory. We believe that at higher $Ra$, where the dissipation progressively goes into the boundary layers, and using longer runs to average out the fluctuations, the small discrepancy will disappear. Using runs A1, A2 and A3, where only $Ra$ varies, we can test the $Ra^{1/4}$ power law predicted in  (\ref{eq:6_16}). The least squares fit to a straight line in $\log (Nu)$ vs $\log (Ra)$ space has a slope of 0.257, rather close to the predicted slope.  

We now look at the larger $\Gamma$ cases, runs A4, A5 and B2, corresponding to the more compressible cases. In the case run A4, at density ratio 5, the predicted and numerical Nusselt numbers are reasonably close, but in the most extreme cases of density ratio 10 the predicted $Nu$ is only 61\% of the numerical value for run B2, and in the case A5 predicted $Nu$ is 71\% of the numerical $Nu$. Part of this discrepancy is down to the boundary layer ratios, which become very large at high $\Gamma$, and so small inaccuracy can affect the Nusselt number significantly. If we use the numerical boundary layer ratios in (\ref{eq:6_16}) rather than the theoretical ones for run B2, the predicted $Nu$-nblr rises to 3.12, but even this is only 76\% of the numerical value, and A5 similarly improves but still is too low. We again conclude that in runs B2 and A5 the assumption that the dissipation occurs dominantly in the boundary layers is suspect (particularly near the top boundary) and that higher $Ra$ is needed before it becomes robustly valid.

We now consider the Reynolds number formula, (\ref{eq:6_17}), though it is convenient to express this in terms of Peclet number $Pe = Re Pr$ . If the table 1 parameter values are inserted into (\ref{eq:6_17}) with the value of $C_{Re}$ quoted there, the values of the Peclet number are consistently a factor of about 5 too small compared with the numerical values of $U_T$ and $U_B$ in table 1. There are a number of reasons for this, but the two most important are (i) the power law dependence of the Peclet number with Rayleigh number in Boussinesq convection is slightly less than the predicted 0.5 (Grossmann \& Lohse 2002), and (ii) our runs are for aspect ratio 2, whereas experiments, and the numerical simulations that simulate them (e.g. Silano \textit{et al.} 2010), use aspect ratios of 1 or less. The prefactor in  (\ref{eq:6_17}) is based on experiments at large $Ra$ which mostly used aspect ratios less than unity. The experiments of Qiu \& Tong (2001), see also figure 1 of Grossmann \& Lohse (2002), using water (with $Pr=5.5$)  in a cylinder of aspect ratio unity found $Re=0.085 Ra^{0.455}$. At the run A1 parameters this formula gives $Pe=251$ consistent with a prefactor of $C_{Re}=0.38$ in (\ref{eq:6_17}), much larger than the Grossmann \& Lohse (2002) value. They found very similar $Re$ prefactors for both high and low $Pr$. If we adopt the same procedure as we did to get the Nusselt number prefactors, and normalise using the Boussinesq runs D1, D2 and D3 we obtain $C_{Re}=0.354$ for $Pr=10$, $C_{Re}=0.400$ at $Pr=1$ and $C_{Re}=0.421$ at $Pr=0.25$. Reassuringly, these are all quite close to the Qiu \& Tong (2001) value of $C_{Re}=0.38$. We therefore use these three values of $C_{Re}$ at the appropriate Prandtl number in all our theory calculations. With these prefactors, the  numerical results for $U_T$ and $U_B$ agree reasonably well with the predicted $Pe_T$-theory and $Pe_B$-theory results. The results have some scatter, which seems to reflect the scatter in our computed $r_U$. If we use the computed boundary layer ratios, rather than the asymptotically predicted ratios, there is less scatter in the comparison between computed and theoretical Peclet numbers, though the theoretical Peclet numbers are 
generally a few percent lower that the computed Peclet numbers. Given that the boundary layers are not very thin, these small discrepancies are not unexpected, and overall the predicted Reynolds numbers are in reasonable agreement with those of our \S6 asymptotic theory.

\section{\label{section8}Conclusions}

The scaling laws for heat flux and Reynolds number at high Rayleigh number convection have been derived from the energy balance and entropy balance equations derived in \S3. These scaling laws are derived in terms of the Rayleigh number, the Prandtl number and the temperature ratio $\Gamma$ which measures the strength of the stratification. In the Boussinesq limit, $\Gamma \to 1$, they reduce to the scaling laws of Grossmann \& Lohse (2000). The existence of the well-mixed entropy state, with the entropy changes being mainly confined to thin boundary layers, makes it possible to estimate the terms in the entropy balance equation, so allowing Nusselt number and Reynolds number relationships to be established. The cases treated are those where the viscous dissipation occurs in the boundary layers, the cases labelled as $I_u$ and $I_l$ by Grossmann \& Lohse (2000), the subscripts referring to the high and low  Prandtl number regimes, and the cases where the viscous dissipation is primarily in the bulk, the cases $II_u$ and $II_l$. A limitation of the theory is that both the entropy boundary layers do have to be thin for the theory to be valid. For the top boundary layer to be thin when the stratification is strong, the Rayleigh number has to be very large, which is numerically difficult, so the range of $\Gamma$ which can be tested both numerically  and asymptotically is quite limited. This condition that the top boundary layer is thin is equivalent to the condition that the boundary layers are incompressible, so that a rather simple relationship holds between temperature and entropy within the boundary layers. The more difficult case where the boundary layers are compressible has not yet been solved in closed form, but it is likely to be significantly different from our solutions.  

A feature of this high Rayleigh number anelastic problem is that the top and bottom boundary layers have a different structure, so to determine the scaling laws, boundary layer ratios for the top and bottom boundary layers have to be established. The three key ratios are those for the boundary layer widths, the boundary layer entropy jumps and the horizontal velocities just outside the boundary layers. In \S5 we proposed formulae based on a simple physical picture for these ratios. We have performed some numerical simulations to test these proposed boundary layer ratios, and within the constraints imposed by the numerics, namely not very high $Ra$, we find broad agreement between the theory and the numerics. Another important assumption for Grossmann-Lohse theory to be valid is the existence of a wind of turbulence. Our numerics suggest that this feature persists in our simulations. There is, however, still some uncertainty about whether the horizontal length scale
of that wind, which controls the boundary layers, remains at the vertical length scale $d$ as the stratification $\Gamma$ increases, or whether it becomes smaller at the top boundary than the bottom boundary at large $\Gamma$.    

We have also tested the theoretically derived Nusselt number and Reynolds number relationships against the numerics, in the case
where the viscous dissipation is mainly in the boundary layers, the only numerically accessible case. Using the prefactors determined in the Boussinesq case, which are the only free parameters in the theory, the Nusselt numbers obtained are in reasonable agreement with the theory, again noting the numerical limitations preventing accurate agreement. A problem was encountered when comparing with the theoretical Reynolds numbers, in that the theory using the original Grossmann-Lohse prefactors gave smaller $Re$ than did the numerics. However, the disagreement seems to be due more to issues with the Boussinesq problem rather than to its extension to the anelastic case, in particular to the difficulty of establishing a single prefactor over a huge range of $Ra$ and to dependence of $Re$ on the aspect ratio. When the prefactors were determined by normalizing on our Boussinesq runs, the issue was resolved.

 We have focussed here on the case of no-slip boundaries, as this seems the simplest case in which scaling laws can be derived from first principles without introducing arbitrary constants into the formulae. There are, however, many similar problems which could be addressed which are of great astrophysical interest: the case of stress-free boundaries is thought to be particularly relevant to stellar convection zones. Even within the context of our simplified no-slip problem, the case of compressible boundary layers would be of interest. We found it most convenient to consider fixed entropy boundary conditions, but other boundary conditions, such as fixed temperature or fixed flux are of interest too. Another issue that could be explored are the differences  between temperature diffusion and entropy diffusion cases. In our particular problem, with incompressible boundary layers, the differences appear to be quite minor, but this is not necessarily the case if more challenging cases are addressed.  Given the growing importance of the anelastic approximation in exploring a very wide range of exciting astrophysical problems, a firmer understanding of the fundamental behaviour of high Rayleigh number anelastic convection would be very valuable.

\begin{acknowledgements}

\noindent{\bf \large Acknowledgements}
This work was partially funded by the STFC grant ST/S00047X/1 held at the University
of Leeds. The partial funding from the subvention of the Ministry of Science and Higher Education in Poland as a part of the statutory activity and the support of the National Science Centre of Poland (grant No. 2017/26/E/ST3/00554) is gratefully acknowledged. 
 The computational work was performed on the ARC clusters, part of the high performance computing facilities at the University of Leeds, and on the COSMA Data Centre system at Durham University, operated on behalf of the STFC DiRAC HPC.

\end{acknowledgements}

\begin{appendices}

\appendix

\section{Form of the anelastic temperature perturbation}\label{appA}

Taking the horizontal average of the anelastic continuity equation (2.2), and using the $u_z=0$ boundary
conditions gives
\begin{equation}
 \langle u_z \rangle_{h} = 0.
\label{eq:appendix_1}
\end{equation} 
Using  (\ref{eq:2_1}) and  (\ref{eq:2_13}), the $z$-component of the  anelastic equation of motion can be written
\begin{equation}
{\bar \rho} \frac{\partial u_z}{\partial t} + \nabla \cdot ({\bar \rho} u_z {\bf u} ) + \nabla p = -g \rho + 
\mu \left( \nabla^2 u_z +\frac{1}{3} \frac{\partial}{\partial z} \nabla \cdot {\bf u} \right).  
\label{eq:appendix_2}
\end{equation}
Taking the horizontal average of (\ref{eq:appendix_2}) we see that, using (\ref{eq:appendix_1}), the viscous term vanishes to leave 
\begin{equation}
\frac{\mathrm{d}}{\mathrm{d}z}\left\langle \bar{\rho}u_{z}^{2}\right\rangle _{h}+
\frac{\mathrm{d}\left\langle p \right\rangle _{h}}{\mathrm{d}z}=
-g\left\langle \rho \right\rangle _{h} =  - g {\bar \rho} \left( \frac{\langle p \rangle_h}{\bar p}  - \frac{\langle T \rangle_h}{\bar T}  \right) \label{eq:appendix_3}
\end{equation}
using (\ref{eq:2_4}a).
%Because the reference state satisfies the hydrostatic equation exactly, we can express
%his equation in terms of the full thermodynamic variables $\hat p = \bar p + \epsilon p$,
%$\hat \rho = \bar \rho + \epsilon \rho$ and $\hat T = \bar T + \epsilon T$, so
%\begin{equation}
%\frac{\mathrm{d}}{\mathrm{d}z}\left\langle \bar{\rho}u_{z}^{2}\right\rangle _{h}+
%\frac{\mathrm{d}\left\langle \hat{p}\right\rangle _{h}}{\mathrm{d}z}=
%-g\left\langle \hat{\rho}\right\rangle _{h}.\label{eq:appendix_4}
%\end{equation}
In the bulk, entropy is well-mixed, so it is constant there so differentiating (\ref{eq:2_4}b) and using (\ref{eq:2_4}a),
\begin{equation}
 R \frac{d}{dz} \left( \frac{ \langle p \rangle_h}{\bar p} \right) = c_p \frac{d}{dz} \left( \frac{ \langle T \rangle_h}{\bar T}\right)
\label{eq:appendix_4}
\end{equation}   
in the bulk. Using the adiabatic reference state hydrostatic and perfect gas equations, with  (\ref{eq:2_8}), this can be written 
\begin{equation}
 \frac{d \langle p \rangle_h}{dz}  = c_p  {\bar \rho} \frac{d \langle T \rangle_h}{dz} - \frac{g {\bar \rho}}{\bar p}  \langle p \rangle_h +
\frac{g {\bar \rho}}{\bar T} \langle T \rangle_h
\label{eq:appendix_5}
\end{equation} 
 and on substituting this into (\ref{eq:appendix_3}) we obtain
%\begin{equation}
%\hat p = K {\hat \rho}^\gamma , \quad \hat p = R \hat \rho \hat T,
%                      \label{eq:appendix_5}
%\end{equation}
%and differentiating these with respect to $z$ gives
%\begin{equation}
 %\frac{d {\langle \hat T \rangle_{h}}}{dz} = - \frac{g}{c_p} - \frac{1}{c_p {\langle \hat \rho \rangle_{h}}}\frac{d}{dz} \langle \bar \rho u_z^2 \rangle_{h}.
%\label{eq:appendix_6}
%\end{equation}
%We now subtract off the adiabatic reference state part of this equation to get
\begin{equation}
 \frac{d {\langle T \rangle_{h}}}{dz} =  - \frac{1}{c_p {\bar \rho}}\frac{d}{dz} \langle \bar \rho u_z^2 \rangle_{h},
\label{eq:appendix_6}
\end{equation}
which is valid in the bulk. Integrating this across the bulk from $z=\delta^{th}_B$ to $z=d-\delta^{th}_T$, and assuming $u_z$
is negligible close to the boundaries,
\begin{equation}
\langle T_{bulk}(d-\delta^{th}_T) \rangle_{h} - \langle T_{bulk}(\delta^{th}_B)  \rangle_h =  \int_{\delta^{th}_B}^{d-\delta^{th}_T}  
\langle \bar \rho u_z^2 \rangle_{h}  \frac{d}{dz} \left( \frac{1}{c_p \bar \rho} \right) \, dz 
= \Delta T_{vel} > 0,
\label{eq:appendix_7}
\end{equation}
since $\bar \rho$ is monotonic decreasing with $z$. This establishes that in the bulk the gradient $d {\langle T \rangle_h} /dz$ is positive on average, corresponding
to a subadiabatic horizontally averaged temperature gradient. We denote this jump in $T$ across the bulk by $\Delta T_{vel}$ because
it is physically connected to the pressure changes induced by the fluid velocity. A natural question is how large  $\Delta T_{vel}$ is
compared to the jumps in $\langle T \rangle_h$ across the boundary layers, $\Delta T_B$ and $\Delta T_T$. Formally they are both of same order of magnitude
in the anelastic approximation, but $\Delta T_{vel}$ will be small if we are close to Boussinesq or if the Rayleigh number is small.
Numerical evidence is sparse, but figure 4 from Verhoeven et al. (2015) suggests that for their parameters, $Ra=10^6$, $\rho_B/\rho_T = 2.72$ and $Pr=0.7$, their $\Delta T_{vel}$ was small.

\subsection{Positivity of the temperature offsets}

We now consider the temperature offsets at the bottom and top boundaries, $\left\langle T \right\rangle _{h,T}$
and $\left\langle T \right\rangle _{h,B}$. Without numerical simulations, we cannot determine their magnitude,
but we can show that they must both be positive, a useful check on future simulations.

By examining the sum of the temperature jumps across the layer in Figure 1b we can see that
\begin{equation}
\left\langle T\right\rangle _{h,T}-\left\langle T\right\rangle _{h,B}
+ \Delta T_{B}+ \Delta T_{T}= \Delta T_{vel}.
\label{eq:appendix_8}
\end{equation}
Using the incompressible boundary layer forms for the temperature jumps across the boundary layers, (\ref{eq:2_25}),
and the formulae for the ratios of these jumps,
\begin{equation}
r_s = \frac{\Delta S_T}{\Delta S_B}, \quad r_T = \frac{\Delta T_T}{\Delta T_B}, \quad r_s = \Gamma r_T,
\label{eq:appendix_9}
\end{equation} 
equation  (\ref{eq:appendix_8})  becomes 
\begin{equation}
\left\langle T\right\rangle _{h,T}-\left\langle T\right\rangle _{h,B}
+ \frac{T_B \Delta S}{c_p} \frac{(1+r_T)}{(1+ \Gamma r_T)}   = \Delta T_{vel}.
\label{eq:appendix_10}
\end{equation}
A second equation for the temperature offsets can be derived from the boundary conditions. From
(\ref{eq:2_4}a) and (\ref{eq:2_4}b) we can deduce
\begin{equation}
s = \frac{c_p T}{\bar T} - \frac{p}{\bar \rho \bar T} 
\label{eq:appendix_11}
\end{equation}
At $z=0$ and $z=d$ this gives
\begin{equation}
\Delta S = \frac{c_p \left\langle T\right\rangle _{h,B}}{ T_B} - \frac{\left\langle p\right\rangle _{h,B}}{\rho_B  T_B}, \quad 
0 = \frac{c_p \left\langle T\right\rangle _{h,T}}{ T_T} - \frac{\left\langle p\right\rangle _{h,T}}{\rho_T  T_T}.
\label{eq:appendix_12}
\end{equation}
We now use the mass conservation equation (\ref{eq:2_28}) to set the pressure perturbations on the top and bottom boundary equal,
giving
\begin{equation}
\frac{T_B \Delta S}{c_p} =  \left\langle T\right\rangle _{h,B} -  \left\langle T\right\rangle _{h,T} \frac{\rho_T}{\rho_B }.
\label{eq:appendix_13}
\end{equation}
Equations  (\ref{eq:appendix_10})  and  (\ref{eq:appendix_13})  are two equations for the temperature offsets $\left\langle T\right\rangle _{h,B}$ and $ \left\langle T\right\rangle _{h,T}$,
and using $\rho_B/\rho_T = \Gamma^{m}$ the solutions are 
\begin{equation}
 \left\langle T\right\rangle _{h,B} =  \frac{\Delta T_{vel}}{\Gamma^{m}-1} 
+ \frac{T_B \Delta S}{c_p} \left\{ \frac{\Gamma^{m} (1+\Gamma r_T) - (1+r_T)}{(\Gamma^{m} - 1)(1+ \Gamma r_T)} \right\}, 
\label{eq:appendix_14}
\end{equation}
\begin{equation}
\left\langle T\right\rangle _{h,T} =  \frac{\Gamma^{m} \Delta T_{vel}}{\Gamma^{m}-1}
+ \frac{T_B \Delta S}{c_p} \left\{ \frac{(\Gamma-1) \Gamma^{m} r_T}{(\Gamma^{m} - 1)(1+ \Gamma r_T)} \right\}.
\label{eq:appendix_15}
\end{equation}
Since $\Gamma > 1$ and $\Delta T_{vel} > 0$ it follows that both quantities are positive whatever $\Delta T_{vel}$ is.
It is not possible to decide which offset is larger without having more information about  
$\Delta T_{vel}$, but these results confirm that Figure 1b is a plausible sketch of the temperature perturbation,
and will be helpful in testing numerical simulations.

%%%%%%%%%%%%%%%
%Appendix B
%%%%%%%%%%%%%

\section{The case when the dissipation in the bulk dominates the dissipation in the boundary layers}\label{appB}

Grossmann \& Lohse (2000) point out that at low $Pr$ and large $Ra$ it is possible
for the viscous dissipation in the bulk to be larger than the viscous dissipation in the boundary layers. When this occurs, our arguments about the boundary layer ratios in \S5 and the scaling laws in \S6 need revising. We now consider this scenario.

\subsection{The boundary layer ratios}

When the viscous dissipation is mainly in the bulk, equations (\ref{eq:5_1}-\ref{eq:5_7}) still hold, but the argument  for equation (\ref{eq:5_11}) breaks down
because the entropy flux is no longer approximately constant in the bulk because viscous dissipation in the bulk is no longer negligible. 
We can however use the energy flux equation (\ref{eq:3_2}) because when the viscous dissipation is
mainly in the bulk, the work done by buoyancy must balance the viscous dissipation in the bulk, since now the viscous dissipation
in the boundary layers is negligible. 

So
 \begin{eqnarray}
\frac{g}{c_p}\int_{bulk} \left\langle \bar{\rho}u_{z}s\right\rangle _{h}\mathrm{d}z=
\mu\int_{bulk}\left\langle q\right\rangle \mathrm{d}z ,
\label{eq:B_1}
 \end{eqnarray} 
 and since thermal diffusion and the last two terms in (\ref{eq:3_2}) are negligible in the bulk when the boundary layers are thin, 
it follows that $\langle {\bar \rho}{\bar T} u_z s \rangle_h$ will be approximately the same just outside the two 
boundary layers at  $z=\delta_B^{\nu}$ and  $z=d - \delta_T^{\nu}$, so 
\begin{eqnarray}
\rho_B  T_B \langle u_z s \rangle_h \vert_{z=\delta_B^{\nu}} \approx  \rho_T T_T \langle u_z s \rangle_h \vert_{z=d -\delta_T^{\nu}}. 
\label{eq:B_2}
 \end{eqnarray} 
Note this is different from the case where the dissipation was mainly in the boundary layers, when $ \langle {\bar \rho} u_z s \rangle_h$ is approximately constant. As we did in \S5,we horizontally average the dot product of $\bf u$ and (\ref{eq:2_1}), and apply it just outside the boundary layers, at $z=\delta_B^{\nu}$ and  $z=d - \delta_T^{\nu}$. Here we are justified in neglecting the pressure term as we did in \S5, and we also neglect the viscous term. This is not obvious when most of the viscous dissipation is in the bulk, but following Grossmann \& Lohse (2000), we envisage a turbulent cascade, where the dissipation at larger scales is dominated by the inertial term. We therefore adopt  
\begin{equation}
\frac{1}{2}\frac{\partial}{\partial z}\left(\bar{\rho}\left\langle u_{z}u^{2}\right\rangle _{h}\right)
\approx  -\frac{\partial}{\partial z} \langle u_z p \rangle_h + \frac{g}{c_{p}}\left\langle {\bar \rho} u_{z}s\right\rangle _{h} \approx \frac{g}{c_{p}}\left\langle {\bar \rho} u_{z}s\right\rangle _{h}
\label{eq:B_3}
\end{equation}
at $z=\delta_B^{\nu}$ and  $z=d - \delta_T^{\nu}$. Then since in the bulk we expect all velocity components to be of similar magnitude in the bulk, using (\ref{eq:B_2})
\begin{eqnarray}
\rho_B  T_B \frac{U_B^3}{H_B} \approx \rho_T  T_T \frac{U_T^3}{H_T} \Rightarrow r_u \sim \Gamma^{\frac{m}{3}} ,
\label{B_4}
 \end{eqnarray} 
where the pressure scale heights are defined in (\ref{eq:5_10}). This result differs from 
(\ref{eq:5_11}), where the dissipation is in the boundary layers, so that now the horizontal velocity at the top is expected to be considerably faster than the velocity at the bottom, whereas (\ref{eq:5_13})
predicts only a weak dependence on $\Gamma$.

\subsection{The scaling laws when dissipation is in the bulk}

We still expect thin boundary layers even when the dissipation is mainly in the bulk, so (\ref{eq:6_1}),
\begin{equation}
 \frac{F^{super} \Delta {\bar T}}{T_B T_T}  \sim
   \int_0^d \frac{\mu}{\bar T} \left\langle q \right\rangle_h \, dz,  
\label{eq:B_5}
\end{equation}
still applies, but unlike the boundary layer dissipation case, we do not know how the dissipation is distributed over the interior. We therefore assume that the dissipation in the interior can be written as 
$\langle q \rangle_h \sim U_H^3/ 2 H$ where $U_H(z)$ is the horizontally averaged horizontal velocity and $H$ is the local pressure scale height. We don't know how $U_H (z)$ is distributed in $z$, but we argued in \S B1 above that $\rho U_H^3$
is approximately the same at the edge of both boundary layers, so a reasonable assumption for the purposes of estimation
is that 
\begin{equation}
 \rho U_H(z)^3 \sim  \ \textrm{constant} \ \approx \rho_B U_B^3 \approx \rho_T U_T^3.
\label{eq:B_6}
\end{equation}
The form of $U_H(z)$ from our numerical results suggests this might overestimate the dissipation integrated over the whole layer, but nevertheless we adopt (\ref{eq:B_6}) for the rest of this section. Equation  (\ref{eq:B_5}) then becomes
\begin{equation}
 \frac{F^{super} \Delta {\bar T}}{T_B T_T}  \sim 
\int_0^d \frac{\bar \rho U_H^3}{2 \bar T  H} \, dz =
\int_0^d \frac{\rho_B U_B^3 (m+1) \Delta {\bar T}}{2 d {\bar T}^2} \, dz = \frac{\rho_B U_B^3 (m+1)(\Gamma-1)}{2 T_b }. 
\label{eq:B_7}
\end{equation}
From (\ref{eq:2_15}) $F^{super} = Nu k T_B/d$, and writing $U_B$ in terms of the bottom Reynolds number
using  (\ref{eq:6_6})
\begin{equation}
 \frac{Nu k \Delta {\bar T}}{d T_T}   \sim 
 \frac{\mu^3 Re_B^3 (m+1)(\Gamma-1)}{2 T_b \rho_B^2 d^2 }. 
\label{eq:B_8}
\end{equation}
Combining (\ref{eq:2_12}) and (\ref{eq:2_16}), we can write the Rayleigh number as
 \begin{equation}
 Ra = \frac{c_p^2 \Delta {\bar T} d^2 \rho_B^2 \Gamma \ln \Gamma}{\mu k (\Gamma-1)} ,
\label{eq:B_9}
\end{equation}
and combining this with  (\ref{eq:B_8}) and using the definition of the Prandtl number  (\ref{eq:4_1}) we obtain
 \begin{equation}
 \frac{Nu Ra}{Pr^2} \sim \frac{(m+1)\ln \Gamma}{2} Re_B^3,
\label{eq:B_10}
\end{equation}
which is the entropy balance equation in the case where the dissipation is mainly in the bulk rather than the boundary layers. We now use the same boundary layer balance equation as before, but since we expect bulk dissipation only to dominate at low $Pr$, we only use (\ref{eq:4_5}) for the boundary layer ratio, so (\ref{eq:6_11}) becomes
 \begin{equation}
\left( Re_B Pr \right)^{1/2}  =  \frac{Nu (\Gamma - 1)(1+r_s)}{\Gamma \ln \Gamma}.
\label{eq:B_11}
\end{equation}
Combining  (\ref{eq:B_10}) and  (\ref{eq:B_11}) we get the Nusselt number in terms of the Rayleigh number in this
case,
\begin{equation}
 Nu \sim Ra^{1/5} Pr^{1/5} \left(  \frac{2}{m+1} \right)^{1/5}  \frac{ \Gamma^{6/5} \ln \Gamma}{ (\Gamma-1)^{6/5}} 
(1+r_s)^{-6/5}.
\label{eq:B_12}
\end{equation}

\end{appendices}

\end{document}